\documentclass[useAMS,usenatbib,aas_macros]{mn2e}

\input{psfig}
\newcommand{\equnp}[1]{eq.~\ref{eq:#1}}
\newcommand{\se}[1]{\S\ref{sec:#1}}
\newcommand{\fig}[1]{Fig.~\ref{fig:#1}}
\newcommand{\Fig}[1]{Figure~\ref{fig:#1}}
\newcommand{\be}{\begin{equation}}
\newcommand{\ee}{\end{equation}}

\def\no{\noindent}

\newcommand{\msun}{M_\odot}

\newcommand{\ifm}[1]{\relax\ifmmode#1\else$\mathsurround=0pt #1$\fi}
\newcommand{\kms}{\ifmmode\,{\rm km}\,{\rm s}^{-1}\else km$\,$s$^{-1}$\fi}

\newcommand{\ltsima}{$\; \buildrel < \over \sim \;$}
\newcommand{\lsim}{\lower.5ex\hbox{\ltsima}}
\newcommand{\gtsima}{$\; \buildrel > \over \sim \;$}
\newcommand{\gsim}{\lower.5ex\hbox{\gtsima}}

\def\msc{M_{{\rm *crit}}}

\def\msh{M_{\rm shock}}

\def\M11{M_{11}}
\def\V100{V_{100}}
\def\R1{R_{Mpc}}
\def\T6{T_6}

\title[Modelling the galaxy bimodality]
{Modelling the Galaxy Bimodality: Shutdown Above a Critical Halo Mass}

\author[A. Cattaneo et al.]
{A.~Cattaneo$^{1,2,3,\star}$, A.~Dekel$^{1,2,\dagger}$, J.~Devriendt$^4$, B.~Guiderdoni$^4$, J.~Blaizot$^5$\\
\\
$^1$Racah Institute of Physics, Hebrew University of Jerusalem, 91904 Jerusalem, Israel\\
$^2$Institut d'Astrophysique, 98 Boulevard Arago, Paris 75014, France\\
$^3$Astrophysikalisches Institut Potsdam, an der Sternwarte 16, 14482 Potsdam, Germany\\
$^4$Centre de Recherche Astronomique de Lyon, 9 Avenue Charles Andr\'e, 69561, St-Genis-Laval Cedex, France\\
$^5$Max-Planck-Institut f\"ur Astrophysik, Karl-Schwarzschild-Str.1, 85740 Garching, Germany\\
$^\star$acattaneo@aip.de, $^\dagger$dekel@phys.huji.ac.il }
\begin{document}

\pagerange{\pageref{firstpage}--\pageref{lastpage}} \pubyear{2005}

\maketitle

\label{firstpage}


\begin{abstract}

We reproduce the blue and red sequences in the observed joint distribution 
of colour and magnitude for galaxies at low and high redshifts using hybrid
N-body/semi-analytic simulations of galaxy formation.  The match of model 
and data is achieved by mimicking the effects of cold flows versus 
shock heating coupled to feedback from active galactic nuclei (AGNs),
as predicted by Dekel \& Birnboim (2006).
After a critical epoch $z \sim 3$, only haloes below a critical 
shock-heating mass $\msh\!\sim\!10^{12}\msun$ enjoy gas supply by 
cold flows and form stars, while cooling and star formation are shut down
abruptly above this mass.  The shock-heated gas is kept hot because being 
dilute it is vulnerable to feedback from energetic sources such as AGNs 
in their self-regulated mode.  The shutdown explains in detail the 
bright-end truncation of the blue sequence at $\sim L_*$, 
the appearance of luminous red-and-dead
galaxies on the red sequence starting already at $z\sim 2$, the colour 
bimodality, its strong dependence on environment density and its correlations 
with morphology and other galaxy properties.  Before $z \sim 2-3$, even haloes 
above the shock-heating mass form stars by cold streams penetrating through 
the hot gas. This explains the bright star-forming galaxies at $z \sim 3-4$, 
the early appearance of massive galaxies on the red sequence, the high 
cosmological star-formation rate at high redshifts and the subsequent low
rate at low redshifts.

\end{abstract}



\section{Introduction}
\label{sec:intro}

Following the early indications for correlations between global galaxy properties
along the Hubble sequence \citep[e.g.][]{hubble26,humason36},
large statistical surveys, such as the Sloan Digital Sky Survey (SDSS)
and the Two Degree Field (2dF),
have revealed the details of a robust bimodality, which divides the galaxy population
into a ``red sequence" and a ``blue sequence".
The bimodality is seen in
(a) the luminosity function \citep{baldry_etal04,bell_etal03},
(b) the colour distribution and the joint distribution of colour
and magnitude or stellar mass \citep{baldry_etal04,balogh_etal04,hogg_etal04},
(c) the stellar age and the star formation rate \citep{kauffmann_etal03b},
(d) the gas-to-stellar mass ratio \citep{kannappan04},
(e) the bulge-to-disc ratio \citep{kauffmann_etal03b},
(f) the environment density \citep{blanton_etal06}.
and (g) the presence of an X-ray halo \citep{helsdon_ponman03,mathews_brighenti03,osmond_ponman04}.

In the colour--magnitude diagram, the red population the and blue population form two 
stripes. In both sequences,  redder galaxies tend to be brighter, but
more so for the blue sequence, which is also broader.
The blue sequence is truncated at $M_r\sim -22$, while the red one
extends to brighter magnitudes.
The division between the two classes of galaxies is associated with 
a critical stellar mass $\msc \sim 3\times 10^{10}M_\odot$.
Galaxies below $\msc$ are typically blue, star-forming spirals and reside in 
the field.
Galaxies above $\msc$ are dominated by red spheroids of old stars and 
live in dense environments.
The faint red galaxies in dense environments define a third population 
\citep{blanton_etal06}.
This indicates that the mass of the host halo, rather than the stellar mass of the galaxy,
determines whether a galaxy is blue or red.

Already at $z\gsim 1$, the most luminous galaxies form a red population 
separated from the blue sequence \citep{bell_etal04,moustakas_etal04},
whereas at $z\sim 2-4$ there is intense star formation in surprisingly 
big galaxies
(Lyman-break galaxies and SCUBA sources: 
\citealp{shapley_etal04,smail_etal02,chapman_etal03,chapman_etal04}).
These observations indicate that star formation in massive galaxies 
happened relatively early and then shut off.

Current models of galaxy formation have difficulties in explaining the
bimodality features seen at the bright end.
There is no natural explanation for the low rate of gas supply indicated by
the low star formation rates \citep{kauffmann_etal03b}
and the low HI fractions \citep{kannappan04} of massive red galaxies
compared to the the high accretion rates and gas fractions inferred 
for blue spiral galaxies.

In the traditional scenario of galaxy formation  
\citep{rees_ostriker77,silk77,white_rees78,blumenthal_etal84,white_frenk91},
the infalling gas is shock-heated to the virial temperature of the dark halo 
near
the virial radius. The inner gas that can cool radiatively on a free fall time scale falls in and forms stars, while the outer gas is held in quasi-static equilibrium until it eventually cools.
The ``cooling radius" equals the virial radius today for haloes of mass
$M_{\rm cool}\simeq 3\times10^{13}(Z/Z_\odot) f_{\rm b}\,M_\odot$,
where $Z$ is the hot gas metallicity 
and $f_{\rm b}$ is the cosmic baryon fraction.
In more massive halos the supply of cold gas gradually stops.

This  
argument has been successful in explaining the ballpark 
of the mass scale separating galaxies from groups, but when it is applied
to compute galaxy colours, it predicts that massive E and S0 galaxies 
should lie on a continuous extension of the blue sequence rather than 
forming a separate sequence that is systematically redder by 
$u-r\simeq 0.2$ at $M_r\lsim -22$ as seen in the SDSS.
Part of this failure arises from the gradual dependence of cooling on mass.

Analyses in spherical symmetry \citep[][BD03]{birnboim_dekel03} and cosmological
simulations \citep[][K05]{keres_etal05} have shown that in haloes below a 
critical shock-heating mass $\msh \sim 10^{12}\msun$  rapid cooling
does not allow gas compression to support a stable virial shock.
So the gas flows cold and unperturbed into the inner halo.
The actual critical $\msh$ is smaller than the crude cooling mass by a factor
of a few, and is comparable to the halo mass indicated by the observed
bimodality. 
The new picture near the critical scale is of a multiphase 
intergalactic medium (IGM) with a cold phase embedded in a hot medium.

These studies indicate that the transition from efficient build-up by cold
flows to complete dominance of the hot mode 
occurs relatively sharply over a narrow mass range about $\msh$.  
\citet{binney04} and 
\citet[][DB06]{dekel_birnboim06}
noticed that this transition can be sharpened further once shock-heating  
triggers another 
heating meachanism that compensates for subsequent radiative losses. 
Feedback from active galactic nuclei (AGNs) is the leading candidate
for the source of such heating
(\citealp{tabor_binney93, tucker_david97, ciotti_ostriker97, silk_rees98, 
granato_etal04, ruszkowski_etal04,brueggen_etal05,dimatteo_etal05}; 
and references therein).
The role of AGN feedback is supported by 
(a) the ubiquity of supermassive black holes in early-type galaxies 
\citep{magorrian_etal98,vandermarel99,tremaine_etal02},
(b) the evidence from X-ray data that outflows can create deep cavities 
in the hot IGM of galaxy clusters \citep{fabian_etal03,mcnamara_etal05}, 
(c) the estimate that a small fraction of the total energy radiated by 
an AGN over a cosmological time is sufficient,
and (d) the possibility of a radiatively inefficient AGN mode at low 
accretion rates that allows pumping energy into the IGM
without violating the strong constraints on the amount of blue light 
that can be released 
\citep[e.g.][for the particular case of M87]{dimatteo_etal03}.

The critical halo mass for effective feedback
does not seem to have its natural roots in the physics of the AGNs themselves. 
On the other hand, this scale arises naturally as the critical scale
for shock heating. Near the shock-heating scale, the IGM is made of cold 
dense clumps in a hot dilute medium. 
This dilute medium is vulnerable to
AGN feedback. The abrupt shutdown in the accretion of gas by galaxies 
above $\msh$ is  trigerred by 
shock heating and maintained over time by AGN feedback. 

DB06 have studied the potential implications of the interplay between 
the cold/hot flow transition and different feedback sources
and have predicted how this can reproduce the observed bimodality features. 
In this paper, we test this hypothesis in a more quantitative way, by
incorporating the shutdown above a critical mass into simulations of
galaxy formation that combine N-body gravity with semi-analytic modelling (SAM)
of the baryonic processes.
In \se{galics} 
we present the semi-analytic/N-body code.
In \se{failure} 
we show that without the abrupt shutdown of gas accretion above a critical mass
the SAM overpredicts the number of bright blue galaxies and fails to 
reproduce the red colours of massive early-type galaxies.
In \se{cold} 
we summarise the theory of cold flows versus shock heating,
and describe how we implement these two features in the SAM.
In \se{new} 
we explain our picture of the interaction between the massive
black hole and the IGM.
In \se{rresults} 
we find that the addition of such an abrupt shutdown brings the
model into excellent agreement with the data.
In \se{rrobust} 
we demonstrate that the success of this model predominantly 
depends on the requirement that the shock-heated gas never cools and is robust
to the detailed implementation of other physical ingredients in the model.
In \se{disc} 
we interpret the results of our simulations and propose a 
unified scenario for the origin of the blue sequence and the two 
parts of the red sequence.
In \se{conc} 
we summarise our main conclusions.

\section{The Hybrid Semi-analytic/N-body Code} 
\label{sec:galics} 

GalICS (Galaxies In Cosmological Simulations; \citealp{hatton_etal03}) 
is a method that combines high resolution simulations of gravitational 
clustering of the dark matter with a semi-analytic approach to the 
physics of the baryons (gas accretion, galaxy mergers, star formation 
and feedback) to simulate galaxy formation in a 
$\Lambda$CDM Universe. 
The version of GalICS used for this article differs from  
the original version in the star formation law (Eq.~1),
the outflow rate due to supernova feedback (Eq.~2) 
and the cut-off that prevents cooling 
on massive galaxies (Eq.~3).

\subsection{The dark matter simulation}

The cosmological N-body simulation has been 
carried out with the parallel tree code developed by Ninin (1999).
The cosmological model is flat $\Lambda$CDM with
a cosmological constant of $\Omega_{\Lambda}=0.667$, 
a Hubble constant of $H_0=66.7{\rm\,km\,s}^{-1}$,
and a power spectrum normalised to $\sigma_8 =0.88$.
The simulated volume is a cube of $(150{\rm\,Mpc})^3$ with $256^3$ 
particles of $8.3\times 10^9$M$_{\odot}$ each
and a smoothing length of $29.3\,$kpc. 
The simulation produced 100 snapshots spaced logarithmically in the 
expansion factor $(1+z)^{-1}$ from $z=35.59$ to $z=0$.

In each snapshot,
we applied a friend-of-friend algorithm \citep{davis_etal85} to identify 
virialised haloes of more than 20 particles.
The minimum halo mass is thus $1.65\times 10^{11}M_\odot$.
The information about individual haloes extracted from the N-body 
simulation and passed to the 
semi-analytic model is in three parameters: 
the virial mass, the virial density and the spin parameter.
Merger trees are  computed 
by linking haloes identified in each snapshot with their progenitors in the 
one before,
including all predecessors from which a halo has inherited one or more 
particles.
We do not use the substructure provided by the N-body simulation. 
Once a halo becomes a subhalo of another halo, we switch from following 
its gravitational dynamics with the N-body integrator to an approximate 
treatment based on a semi-analytic prescription.          

\subsection{Galaxy formation through the accretion of gas} 

Newly identified haloes are attributed a gas mass by assuming a universal 
baryon fraction of $\Omega_{\rm b}/\Omega_0 = 0.135$. 
All baryons start in a hot phase shock-heated to the virial temperature.
Their density profile is that of a singular isothermal sphere 
truncated at the virial radius.

The cooling time is calculated from the 
density distribution of the hot gas using the cooling function of 
\citet{sutherland_dopita93} .
The gas for which both the cooling time and the free fall time are shorter
than the timestep $\Delta t$ is allowed to cool during that timestep. 
Cooling is accompanied by inflow to maintain the shape of the hot gas 
density profile and is inhibited in haloes 
with positive total energy or with angular momentum parameter $\lambda>0.5$.

A systematic comparison of GalICS and SPH results 
(Cattaneo et al. 2006) 
shows that the galaxy mass function obtained with the GalICS
cooling scheme is very similar to the one found with the SPH method, 
which demonstrates that the cooling algorithm is not an important 
source of error.

\subsection{Morphologies}

GalICS models a galaxy with three components: a disc, a bulge and a starburst. 
Each galaxy component is made of stars and cold gas,
while the hot is considered as a component of the halo.
Only the disc accretes gas from the surrounding halo. 
The bulge grows by mergers and disc instabilities, while the
starburst contains gas in transition from the disc to the bulge.
The disc is assumed to be exponential, with the radius determined by 
conservation of angular momentum. 
The bulge is assumed to have a \citet{Hernquist90} profile and its 
radius is determined based on energy conservation.

Disc instabilities transfer matter from the disc to the bulge until the 
bulge mass is large enough to stabilize the disc \citep{vandenbosch98}. 
Dynamical friction drives galaxies to the centres of dark matter haloes 
and is the most important cause of mergers
(while we also include satellite-satellite encounters). 
The fraction of the disc that transfers to the bulge in a merger grows with
the mass ratio of the merging galaxies from zero, for a very minor,
merger to unity, for an equal mass merger. 
This is of course a simplified picture of the dynamics of morphological 
transformations, but it provides results consistent with essential properties
such as the Faber-Jackson relation and the fundamental plane of spheroids.

\subsection{Star formation and feedback}

The star formation law 
is the same for all components: 
\begin{equation}
\label{eq:sfr} 
\dot{M}_*={M_{\rm cold}\over\beta_*t_{\rm dyn}}(1+z)^{\alpha_*} \ .
\end{equation}
The cold gas mass $M_{\rm cold}$ refers to the component in question,
and $t_{\rm dyn}$ is the dynamical time 
(corresponding to half a rotation 
for a disc and the free fall time for a spheroid).
In the standard GalICS model, 
star formation is activated when the gas surface density is
$\Sigma_{\rm gas}>20\,m_{\rm p}{\rm\,cm}^{-2}$ 
($m_{\rm p}$ is the proton mass).
The star formation efficiency parameter has a fiducial value of
$\beta_*=50$ \citep{guiderdoni_etal98},
which is assumed to be the same at all redshifts, $\alpha_*=0$.  
Gas in transit from the disc to the spheroid passes through a starburst phase 
in which the star formation rate grows by a factor of 10 before 
the gas is converted into bulge stars.
This high star formation rate is obtained 
by assuming that the starburst 
radius is ten times smaller than the final bulge radius 
(see the hydrodynamic simulation in \citealp{cattaneo_etal05}). 
The properties of the galaxy population at low redshifts are insensitive 
to the starbust star formation time scale as long as it is much shorter 
than the Hubble time.

\begin{figure*} 
\noindent
\begin{minipage}{8.6cm}
  \centerline{\hbox{
      \psfig{figure=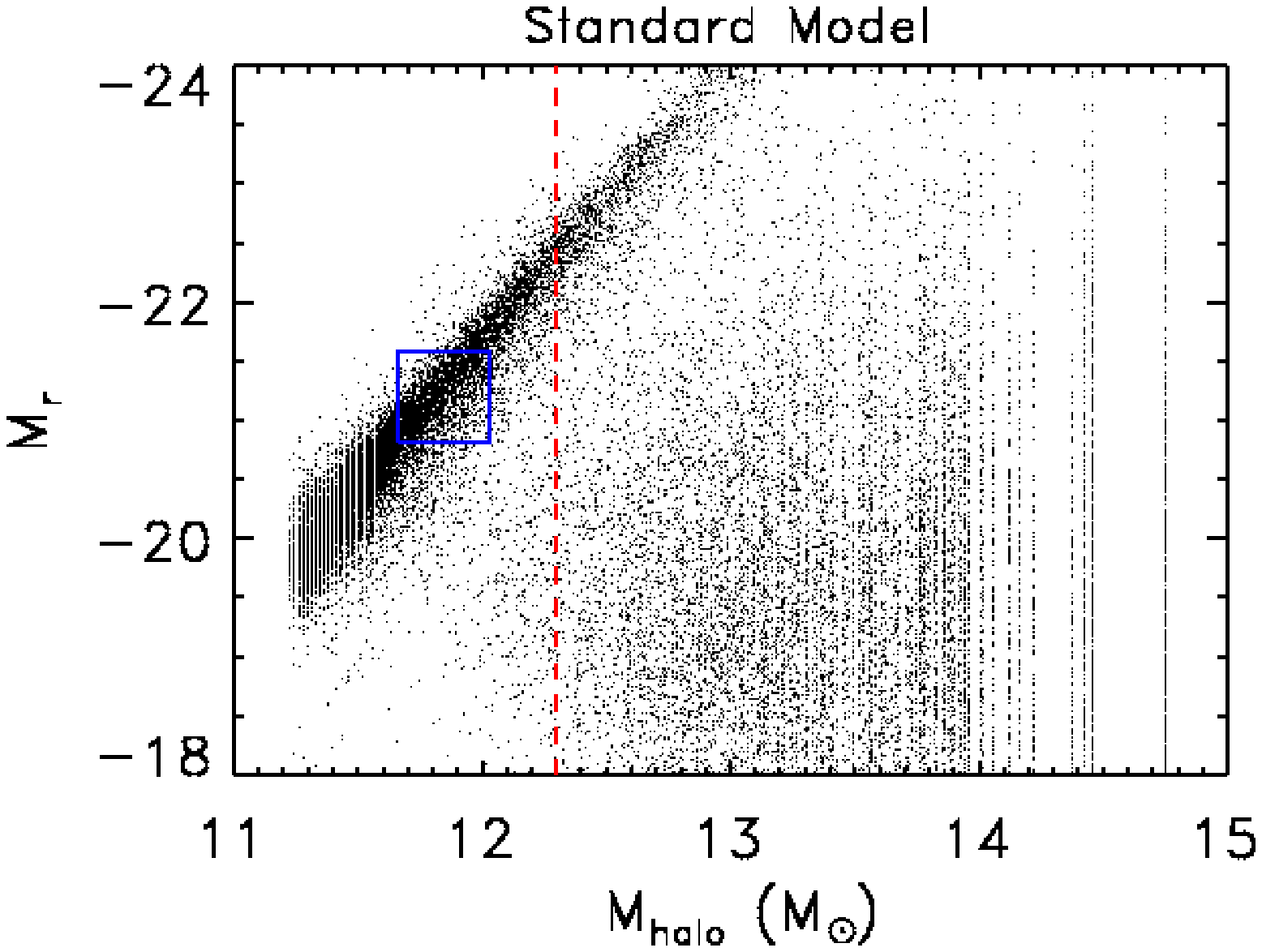,height=5.9cm,angle=0}
  }}
\end{minipage}\    \
\begin{minipage}{8.6cm}
\centerline{\hbox{
\psfig{figure=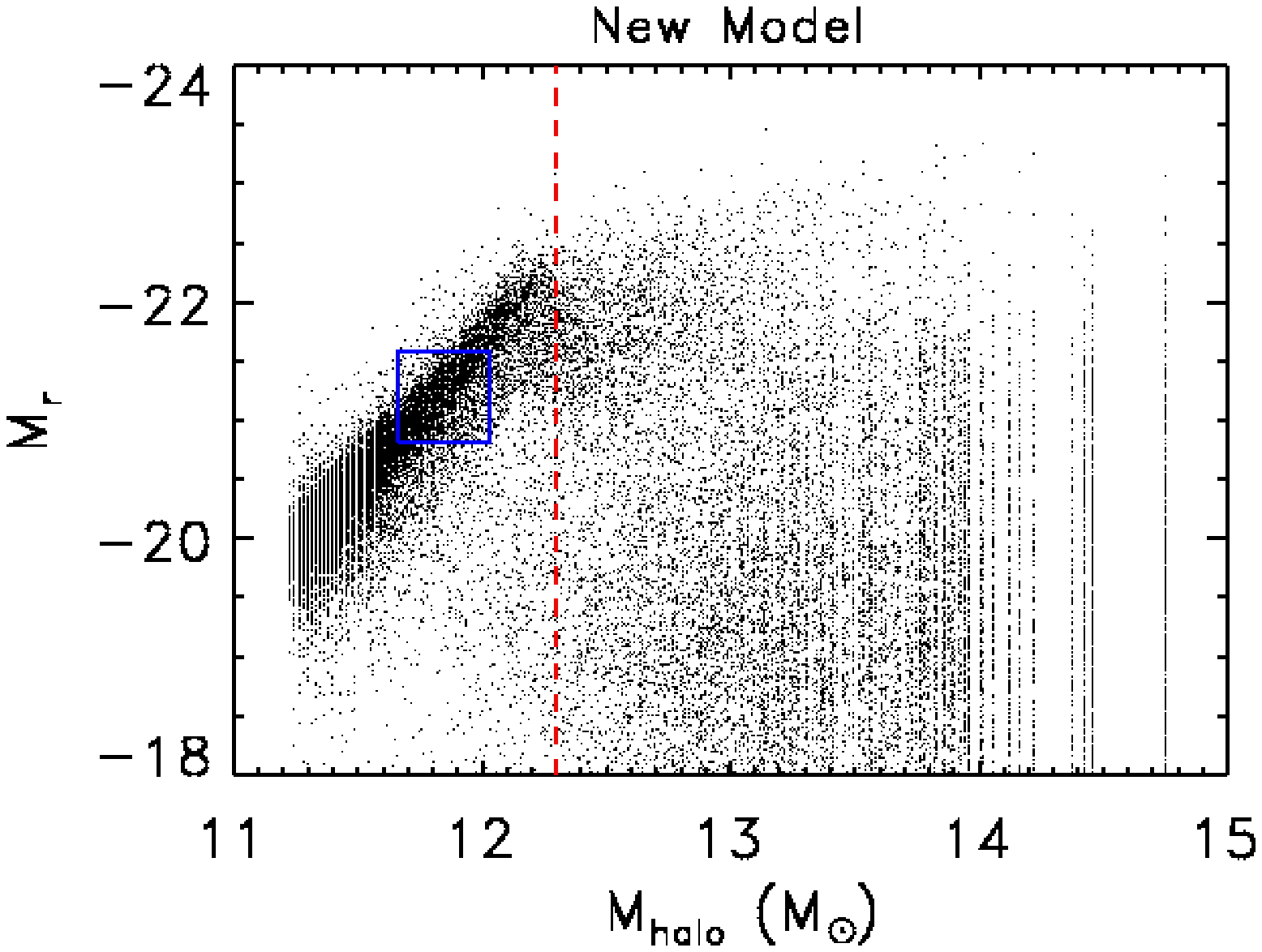,height=5.9cm,angle=0}
  }}
\end{minipage}\    \
\caption{Galaxy luminosity versus halo mass for the ``standard" model (left)
and the ``new" model (right). 
The blue square refers to the Milky Way 
\citep{binney_merrifield98,dehnen_binney98}.
The vertical dashed line marks the shock-heating scale.
The main difference between the two models is in the break in the
relation at $M_{\rm halo}>M_{\rm shock}$ due to the shutdown
above $M_{\rm crit}$ (Eq.~3).  Some differences also show up
below $M_{\rm shock}$ because of the shutdown of cooling
in galaxies where the bulge is the dominant component. 
}
\label{fig:z0_lum_mass}  
\end{figure*}

The mass of newly formed stars is distributed according to the 
\citet{kennicutt83} initial mass function.
Stars are evolved between snapshots using substeps of at most 1\,Myr. 
During each sub-step, stars release mass and energy into the interstellar 
medium. Most of the mass comes from the red giant and 
the asymptotic giant branches of stellar evolution, while most of the energy 
comes from shocks due to supernova explosions. The enriched material 
released in the late stages of stellar evolution is mixed with 
the cold phase, while the energy released from supernovae is used to reheat the cold gas and return it to the hot phase in the halo. 
Reheated gas is ejected from the halo if the potential is shallow enough.
The rate of mass loss through supernova-driven winds $\dot{M}_{\rm w}$ 
is determined by the equation
\begin{equation} 
{1\over2}\dot{M}_{\rm wj}v_{\rm esc}^2
=\epsilon_{\rm SN}\eta_{\rm SN}E_{\rm SN}\dot{M}_*,
\label{eq:mdot} 
\end{equation}
where $E_{\rm SN}=10^{51}\,$erg is the energy of a supernova, 
$\eta_{\rm SN}=0.0093$ is the number of supernovae every
$1\,M_\odot$ of stars and $v_{\rm esc}$ is the escape velocity 
\citep{dekel_silk86}.
In GalICS $v_{\rm esc}\simeq 1.84v_{\rm c}$ for discs and $v_{\rm esc}=2\sigma$ for bulges/starbursts.
The supernova efficiency $\epsilon_{\rm SN}$ is a free parameter, which 
determines the fraction of the supernova
energy used to reheat the cold gas and drive an outflow at the escape speed.
We determine its value by requiring that the model matches the luminosity
and the gas fraction of a galaxy that lives in a Milky Way-type halo.     
This constraint is consistent with the observed normalization 
of the Tully-Fisher relation (\fig{z0_lum_mass}).
Our best fit value $\epsilon_{\rm SN}\simeq 0.2$ agrees with those 
adopted in other SAMs \citep{somerville_primack99,cole_etal00}. 

\subsection{Stardust}

The STARDUST model \citep{devriendt_etal99} gives the spectrum of a stellar 
population as a function of age and metallicity and includes a
phenomenological treatment of the reprocessing of light by dust.
GalICS uses STARDUST to compute the spectrum of each component and 
output galaxy magnitudes.
Dust absorption is computed with a phenomenological extinction law that
depends on the column density of neutral hydrogen, the line of sight 
and the metallicity of the obscuring material.
The reemitted spectrum is the sum of four templates (big and small carbon 
grains, silicates, and polycyclic aromatic hydrocarbons). 
Their relative weights are chosen to reproduce the relation between 
bolometric luminosity and infrared colours observed locally in IRAS galaxies.
In GalICS we consider only the self-absorption of each component.
The mean column density is determined assuming spherical symmetry for 
the starburst and the bulge, while in the disc component there is a 
dependence on a randomly selected inclination angle.

\begin{figure*}
\noindent
\begin{minipage}{8.6cm}
  \centerline{\hbox{
      \psfig{figure=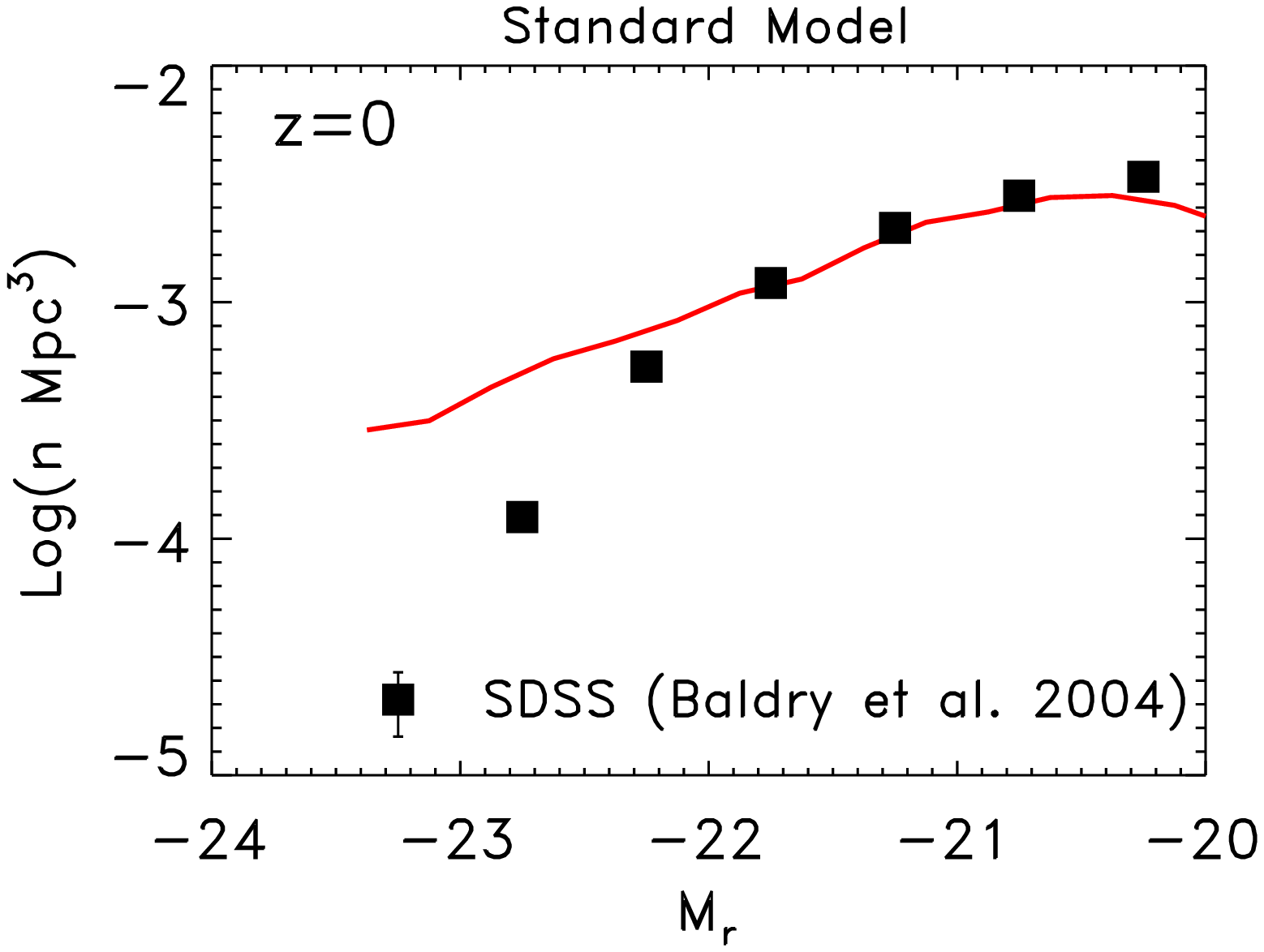,height=5.9cm,angle=0}
  }}
\end{minipage}\    \
\begin{minipage}{8.6cm}
\centerline{\hbox{
\psfig{figure=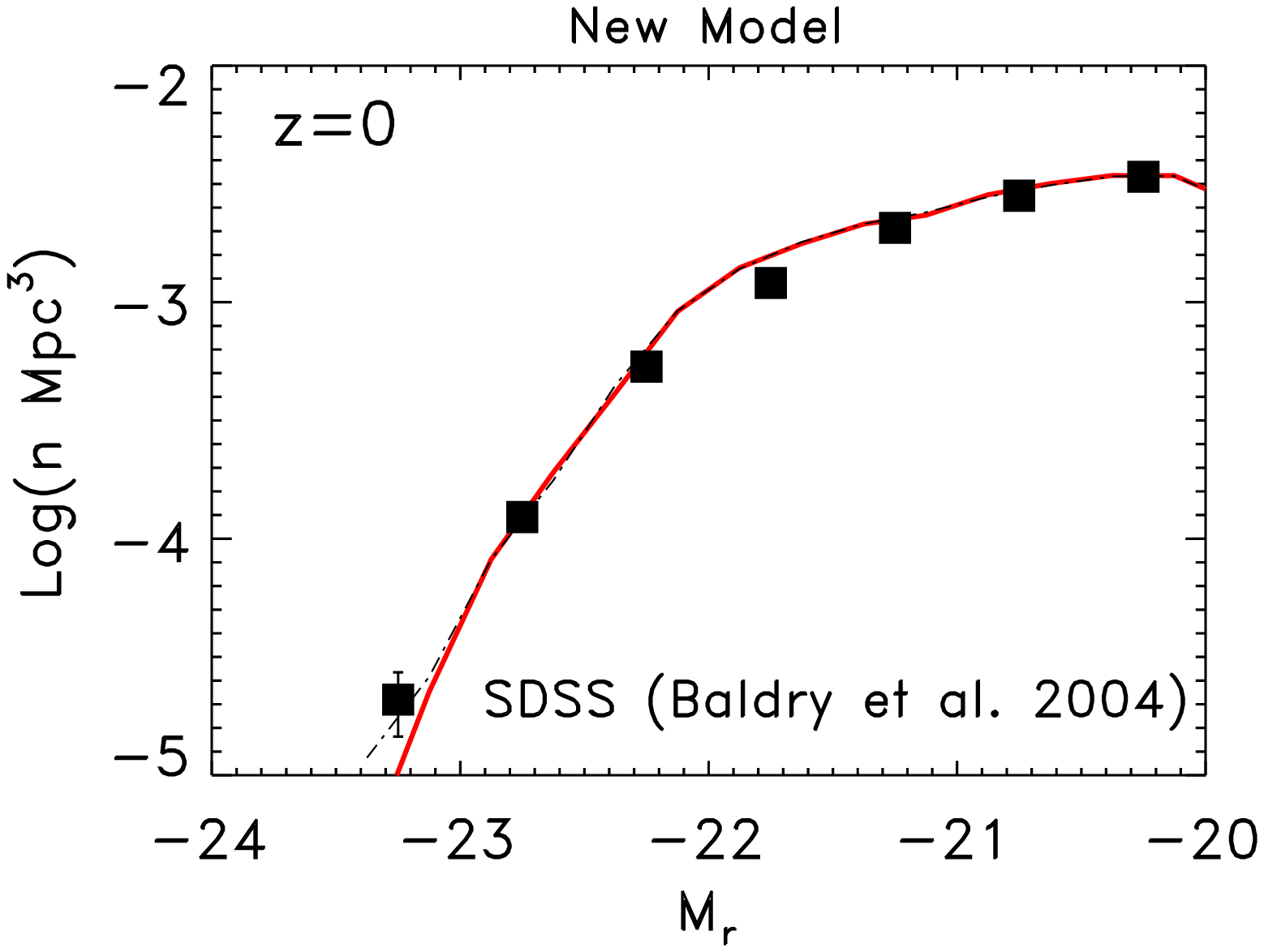,height=5.9cm,angle=0}
  }}
\end{minipage}\    \
\caption{Luminosity function in the $r-$band.
The model predictions (solid line, red) for the ``standard" model (left) and 
for the ``new" model with shutdown in massive haloes (right) 
compared to the observed luminosity function 
(symbols) observed by SDSS \citep{baldry_etal04}.
The dotted-dashed line (right) illustrates the effect of lowering
the critical redshift from $z_{\rm c}=3.2$ to $z_{\rm c}=3$.}
\label{fig:z0_lf} 
\end{figure*}
\begin{figure*}
\noindent
\begin{minipage}{8.6cm}
  \centerline{\hbox{
      \psfig{figure=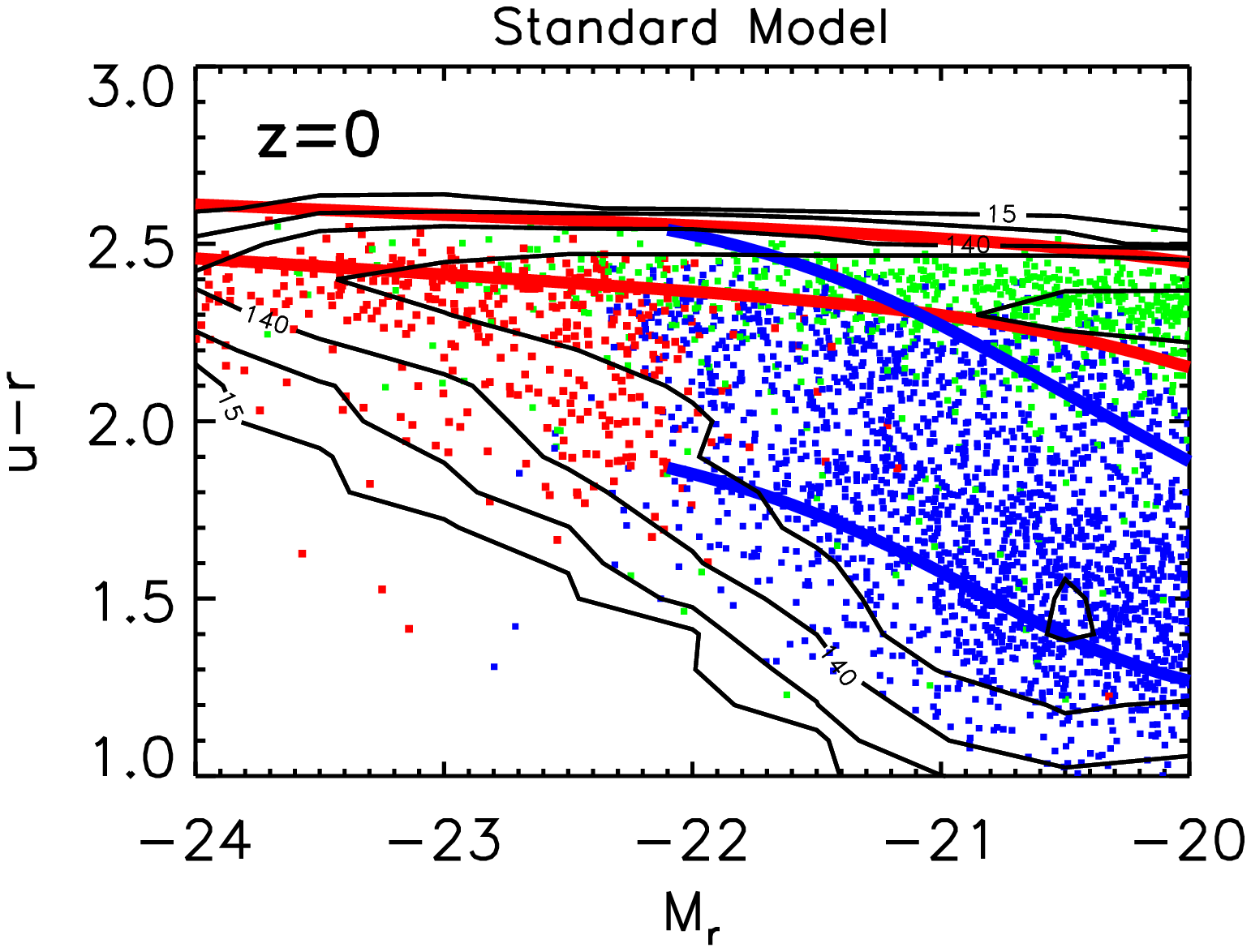,height=5.9cm,angle=0}
  }}
\end{minipage}\    \
\begin{minipage}{8.6cm}
\centerline{\hbox{
\psfig{figure=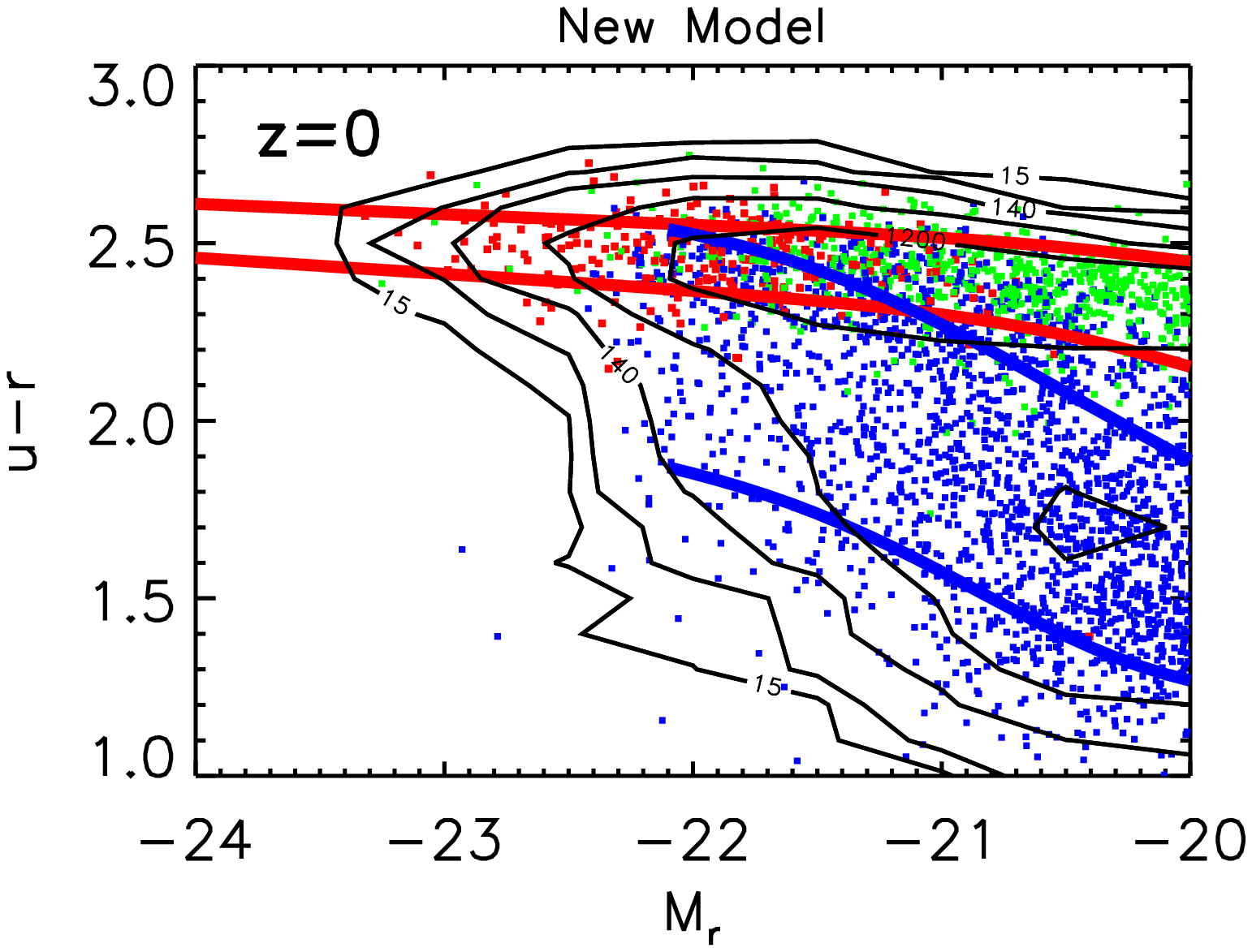,height=5.9cm,angle=0}
  }}
\end{minipage}\    \
\caption{Colour--magnitude diagram at $z=0$ using $u-r$ versus $M_r$ 
(SDSS magnitudes).
Model predictions (dots), ``standard" (left) and ``new" (right), 
compared to the observed loci of the red/blue sequence in the SDSS data 
\citep{baldry_etal04} marked by the pairs of red/blue curves.
Galaxies in haloes below $M_{\rm shock}$ are marked blue; they are 
typically the only galaxy in their halo.
Galaxies in haloes above $M_{\rm shock}$ are marked red if they are the central 
galaxy of their halo and green if they are satellites.
The contours, directly comparable to Fig.~2 of \citet{baldry_etal04}.,
refer to the density of points in units of
${\rm mag}^{-2}{\rm Mpc}^{-3}$) and highlight the bimodality.
}
\label{fig:z0_cm} 
\end{figure*}
\begin{figure*}
\noindent
\begin{minipage}{8.6cm}
  \centerline{\hbox{
      \psfig{figure=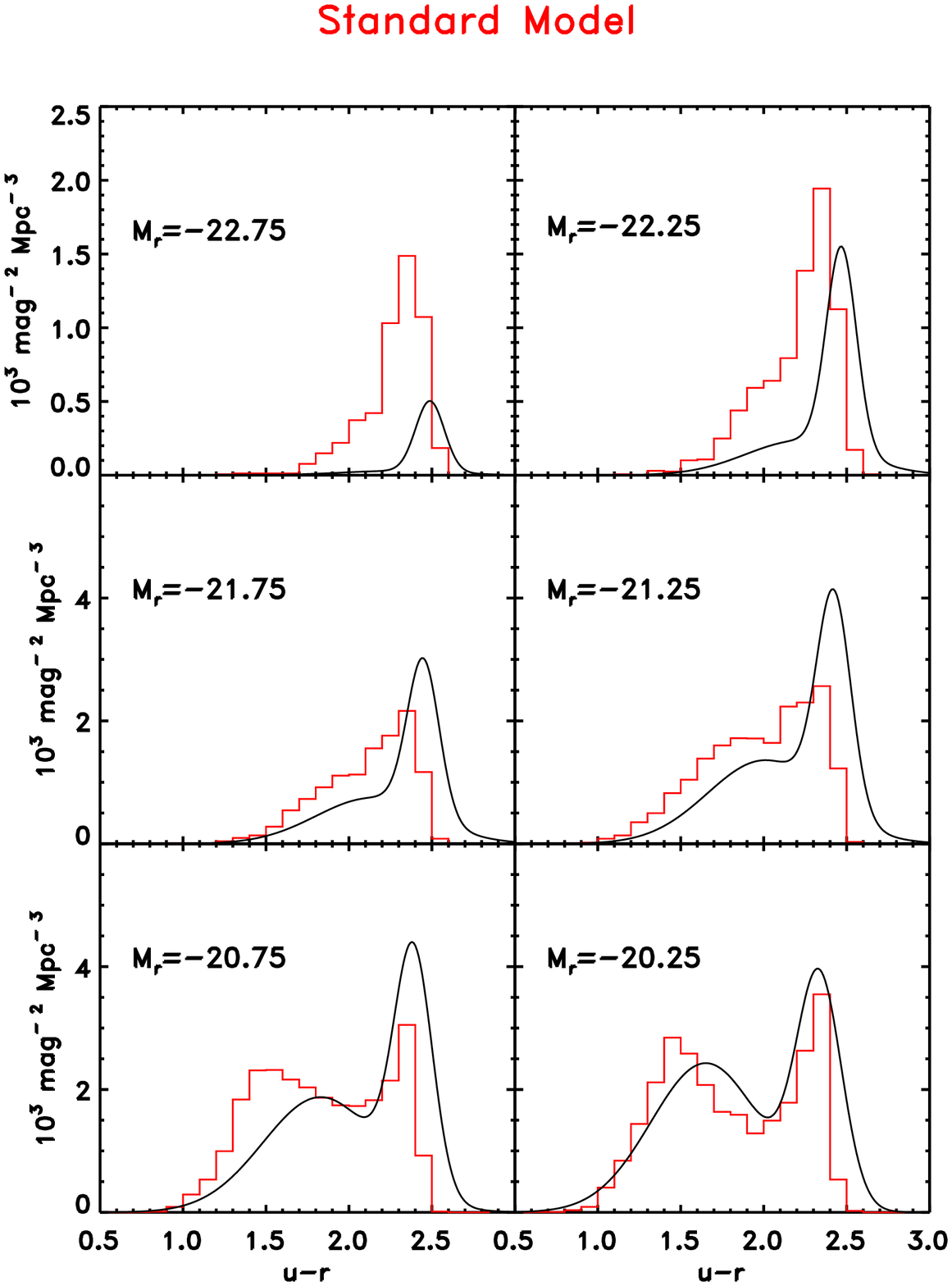,height=13.cm,angle=0}    
  }} 
\end{minipage}\    \
\begin{minipage}{8.6cm}
\centerline{\hbox{
\psfig{figure=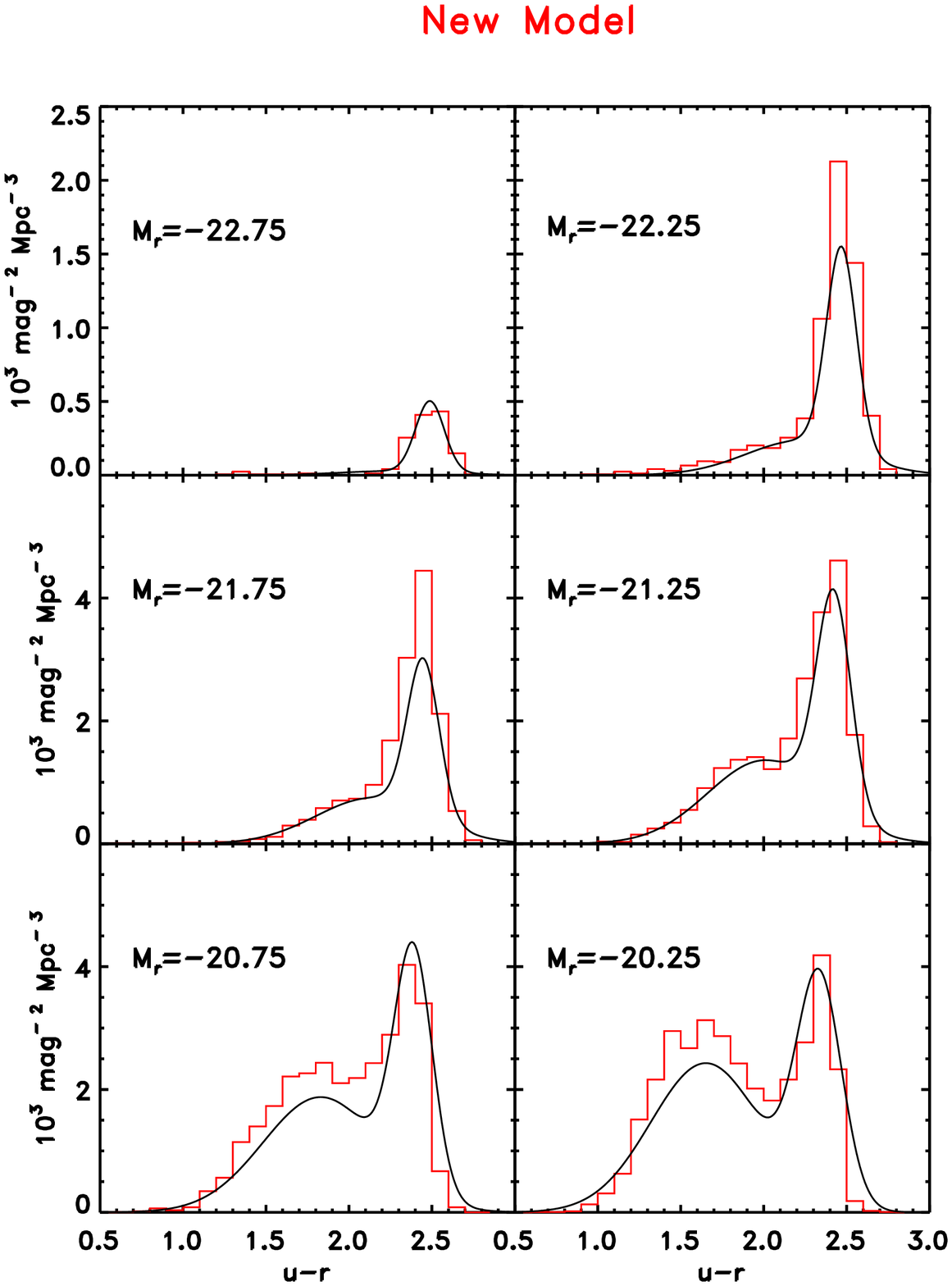,height=13.cm,angle=0}
  }}
\end{minipage}\    \
\vskip 0.5cm
\caption{$u-r$ colour distribution in $r$-band magnitude bins.
The model predictions (red histogram), ``standard" (left) and ``new" (right),
compared to the SDSS data (black smoothed histogram) from \citet{baldry_etal04}.
The ``new" model parameters are $M_{\rm shock}=2\times 10^{12}M_\odot$ 
and $z_{\rm c}=3.2$ (\equnp{mcrit}).
The star formation efficiency increases at high redshift with a power 
of $\alpha_*=0.6$ (\equnp{sfr}) and the energetic efficiency of
supernova feedback is $\epsilon_{\rm SN}=0.2$ (\equnp{mdot}).}
\label{fig:z0_hist} 
\end{figure*}

\section{Failure of the standard model} 
\label{sec:failure}

Semi-analytic models based on the standard scenario (\se{galics})
have difficulties in explaining the early formation of bright elliptical
galaxies followed by passive evolution. 
At $z=0$, the luminosity function shown in \fig{z0_lf} demonstrates
that the standard model predicts too many bright galaxies compared to SDSS.

\Fig{z0_cm} and \fig{z0_hist} illustrate in two different ways that the 
model bright ellipticals are also not red enough. 
In \fig{z0_cm} we see that the brightest galaxies 
(which are also the central galaxies of massive haloes) 
fall below the colour range of red galaxies in the SDSS. 
\Fig{z0_hist} displays the galaxy colour distribution in different 
magnitude bin.  A comparison of the predictions of the standard model 
with the SDSS data in the $-23<M_r<-22.5$ bin shows that
(a) there are many more galaxies in the model than in the data,
(b) the predicted peak of the colour distribution is bluer by $u-r\simeq 0.2$,
and (c) the model distribution is skewed toward blue colours.

At $z\sim 1$, \fig{z1_cm} shows that the model fails to predict the 
existence of bright red galaxies. Instead, the most luminous galaxies 
are predicted to lie on an extension of the blue sequence,
in conflict with the COMBO-17 data \citep{bell_etal04}, which reveal
the bimodality already at $z \sim 1$. 

\fig{z3_lf} shows the failure of the standard model in matching the luminosity
function of Lyman-break galaxies \citep{steidel_etal99} at $z \sim 3$.
With its  star formation rate, the model 
cannot explain the presence of massive star-forming galaxies at $z\sim 3$.
In the whole computational volume, there is only a handful of galaxies with
$M_V<-23$ (\fig{z3_cm}).
The star-formation history shown in \fig{madau} demonstrates again that
the model star formation is not sufficient at high redshifts, 
and that it is too high at low redshifts.
These discrepancies are generic to the standard scenario
and are not specific to the way this scenario is implemented in GalICS.
\cite{blaizot_etal04} were able to use GalICS without the modifications proposed
here and they did find a reasonable agreement with the \citet{steidel_etal99} luminosity function at $z\sim 3$, but they used a non-standard feedback model, which spoils  the fit to joint distribution of colour and magnitude at $z\sim 0$.

\section{Cold Flows versus Shock Heating} 
\label{sec:cold}

A stable extended shock can exist in a spherical system
when the pressure that develops
in the post-shock gas is sufficient to balance the gravitational 
attraction toward the halo centre. 
For adiabatic contraction, this requirement translates into a condition 
on the adiabatic index $\gamma\simeq 5/3$, 
defined as the ratio between the specific 
heats at constant pressure and constant volume.

BD03 have extended this result  
to the case where energy is lost at a rate $q$ 
by replacing the adiabatic index $\gamma$ with the effective polytropic index
$\gamma_{\rm eff}\equiv{\rm d\,ln}p/{\rm d\,ln}\rho
=\gamma-(\rho/\dot{\rho})(q/e)$, where $e$ is the  
internal energy. This can be rewritten as 
$\gamma_{\rm eff}=\gamma-t_{\rm comp}/t_{\rm cool}$, where 
$t_{\rm cool}\equiv e/q$ is the cooling time of the post-shock gas
and $t_{\rm comp}\equiv\rho/\dot{\rho}=-(\nabla v)^{-1}=-r_{\rm s}/(3u_1)$ 
is its compression time, with $r_{\rm s}$ the shock radius and $u_1$
the post-shock radial velocity.
The first equality is the continuity equation
and the second assumes that the post-shock radial flow pattern is homologous.
Using perturbation analysis, BD03 showed that the criterion for a stable
shock is $\gamma_{\rm eff} >10/7$ (for a monoatomic gas in spherical
symmetry), which translates to $t_{\rm cool}^{-1} < t_{\rm comp}^{-1}$.
This defines a critical halo mass, above which the gas is shock-heated near
the virial radius, and below which the gas flows cold and unperturbed into the
inner halo, where it may eventually shock.
This result has been confirmed by BD03 using spherical hydrodynamical 
simulations.
A similar conclusion was obtained earlier in one-dimensional simulations
of pancake formation \citep{binney77}. 
 
DB06 have applied this theory in a cosmological
context. They have assumed that the pre-shock gas traces the
dark matter distribution in its free fall, and   
the strong shock condition determines the post-shock infall speed, 
density and temperature.
They infer a shock-heating scale of 
$M_{\rm shock}\sim 6\times 10^{11}M_\odot$ for a shock
in the inner halo ($r\sim 0.1r_{\rm vir}$),
and $M_{\rm shock}\sim 3\times 10^{12}M_\odot$ for a shock at the 
virial radius.

Cosmological simulations with smoothed particle hydrodynamics 
(\citealp{fardal_etal01}; K05) and with an adaptive Eulerian mesh 
(DB06) have shown that the phenomenon is not an artifact of
the spherical or planar symmetry.
Fig.~6 of K05, based on a simulation with zero metallicity,
shows that the infall is almost all cold at 
$M_{\rm halo} \leq 10^{11}M_\odot$, drops to 
$50\%$ at $M_{\rm shock}\simeq 3\times10^{11}M_\odot$,
and for $z\leq 2$  is almost all hot at $M_{\rm halo} \geq 10^{12}M_\odot$.
This is an underestimate of the critical mass by a factor of order two
because of the assumed zero metallicity.
At higher redshifts the cold fraction does not vanish even above the critical 
mass. At $z=3$ it is $\sim 40\%$ even in the most massive halo contained 
in the computational box, $\gsim 10^{13}M_\odot$.

We model the complex multiphase behavior in an idealised way:
a sharp transition from 100\% cold flow to a complete shutdown
of the accretion at a critical halo mass $M_{\rm crit}(z)$.
We parameterise the critical mass by
\begin{equation}
M_{\rm crit}=M_{\rm shock}\times {\rm min}\{1,\,10^{1.3(z-z_{\rm c})}\} \ .
\label{eq:mcrit} 
\end{equation}
At redshifts below $z_{\rm c}$ the critical mass is the shock-heating
mass, which is expected to be of the order of $10^{12}M_\odot$. 
We treat its exact value as a free parameter determined by the 
best fit to different constraints.
The predicted presence of cold flows above the shock-heating mass at 
$z>z_{\rm c}$ is mimicked by the exponential increase in $M_{\rm crit}$.
The critical redshift $z_{\rm c}$, predicted in the ballpark of $\sim 2-3$,
is also adjusted for best fit.

\section{New elements in the galaxy formation scenario}
\label{sec:new}

\subsection{shutdown by virial shock heating and AGN feedback} 

The critical stellar mass of $3\times 10^{10}\msun$ associated with the
observed
bimodality corresponds to a critical halo mass of $\lsim 10^{12}M_\odot$. 
Star forming galaxies are mainly confined to haloes below this mass 
and most galaxies in haloes above this threshold are evolving passively 
\citep{kauffmann_etal04}.
The main reason for the failure of the standard model at low redshift is 
the cooling of hot gas onto the central galaxies of haloes that are more 
massive than the critical scale.
The traditional cooling scale \citep{rees_ostriker77} is somewhat too big
and is not associated with a transition that is sharp enough to explain
the sharp colour bimodality and the steep drop of the luminosity 
function beyond $L_*$.

The gas that cools and accumulates in the central galaxies of massive haloes
will inevitably form stars unless there is a mechanism that reheats it or 
removes it.
AGNs provide the most plausible source of energy for the job 
\citep{binney_tabor95,tucker_david97,ciotti_ostriker97,silk_rees98,
nulsen_fabian00,springel_etal05}.
The supporting evidence comes from several lines of argument such as 
(a) the abundance of supermassive black holes in massive spheroids
together with
the correlation between black hole mass and bulge mass
\citep{marconi_hunt03,haering_rix04} or the
bulge's velocity dispersion
\citep{merritt_ferrarese01,tremaine_etal02},
(b) the concordance between the quasar epoch and the stellar ages of 
early-type galaxies \citep{granato_etal01,cattaneo_bernardi03},
(c) the connection of black hole growth and elliptical galaxies with mergers 
\citep{toomre_toomre72,stockton99},
and (d) the observation of powerful outflows from active galaxies 
\citep[e.g.,][]{mcnamara_etal05}.

Supernova feedback is not powerful enough to affect the gas 
in massive haloes significantly unless the rate of formation of massive stars 
is especially high, but then these galaxies would no longer be red.
A similar argument can apply to feedback from quasars since
since early-type galaxies are $10^4$ times more common than quasars 
at low redshifts \citep{wisotzki_etal01}.
This is a serious problem for models in which AGNs are supposed to
interact with the halo gas  
through radiative processes such as radiation pressure or Compton scattering.
However, there is evidence showing that AGN outflows are possible not only 
during the main phase of black hole growth, which was very brief and 
happened at high redshift, but also when the black hole accretion rate is 
so low that the AGN is not optically luminous (see e.g. the adiabatic 
inflow-outflow solution by \citealp{blandford_begelman99} and
the study of the jet in M87 by \citealp{dimatteo_etal03}).  
In the inflow-outflow model, the density of the accretion flow is  
too low for the gas to radiate efficiently. 
The accretion power is released mechanically, through particle jets, 
rather than radiatively, through the emission of optical/UV light.
In this scenario, one possible way for the AGN to heat the gas is 
through shocks produced by the propagation of jets 
(e.g. \citealp{reynolds_etal01,fabian_etal03,omma_etal04}),
although the problem of how the heat generated by this process 
is distributed in the hot gas remains open
\citep[see also][and references therein]{begelman_nath05,fabian_etal05}. 

One difficulty in connecting AGN feedback to the galaxy bimodality is
that the critical scale does not seem to affect black hole growth.
For example, the correlation between black holes and bulges extends down to  
small bulges as in the Milky Way \citep{marconi_hunt03,haering_rix04},
while cooling seems to shut down only in massive galaxies. 

Our proposal here (and in DB06) is that AGN feedback is switched on by the 
change in the large scale black-hole environment occuring once the gas 
is shock heated above the shock-heating scale.
While the cold phase is fragmented in dense clouds, the hot phase is 
distributed much more uniformly in a low density medium.
Outflows from the AGN will preferentially propagate through the less 
resistive hot gas that fills the space between the clouds and leave the 
cold clouds behind instead of blowing them away 
\citep[e.g.,][in the case of supernova blast waves.]{mckee_ostriker77}.
The dilute hot gas is thus more vulnerable to the propagation of shock fronts.
This serves as the basis for the new scenario simulated in the current paper.

We thus adopt the assumption that
all massive galaxies contain a supermassive black hole, which can accrete 
gas and deposit energy in the surrounding gas,
but the capacity of the gas to react to this injection depends on the 
presence of a  shock-heated component.
As soon as this condition is fulfilled, the gas begins to expand and the 
black hole accretion rate drops until it stabilises at a value that
over time compensates the radiative losses of the hot gas.
The onset of this self-regulated accretion cycle prevents the 
shock-heated gas from ever cooling.
We model this physical scenario by interpreting the critical mass 
that separates the regimes of cold and hot accretion (\equnp{mcrit}) as
a sharp cooling cut-off. 

When most of the mass is in the hot phase, the cold clouds also become 
vulnerable to feedback. We therefore assume that when 
$M_{\rm halo}>M_{\rm crit}$ the cold gas in the central galaxy is 
reheated to the virial temperature of the halo.
Black hole and supernova winds from gas-rich galaxy mergers will also sharpen
the transition to a hot IGM because the cold streams in free fall 
will go through a shock when they hit the winds blown from the galaxy.
We account for this scenario by preventing cold flows on galaxies in which 
the bulge is the main component as a massive bulge correlates with a 
massive black hole and, therefore, with a history of high AGN activity.
This bulge-related condition improves the details of the quantitative 
agreement with the SDSS colour--magnitude distribution, but 
we will see below 
(\se{rrobust}) that it plays only a minor role in curing the discrepancies
and achieving a good match with the observed features.

\begin{figure*}
\noindent
\begin{minipage}{8.6cm}
  \centerline{\hbox{
      \psfig{figure=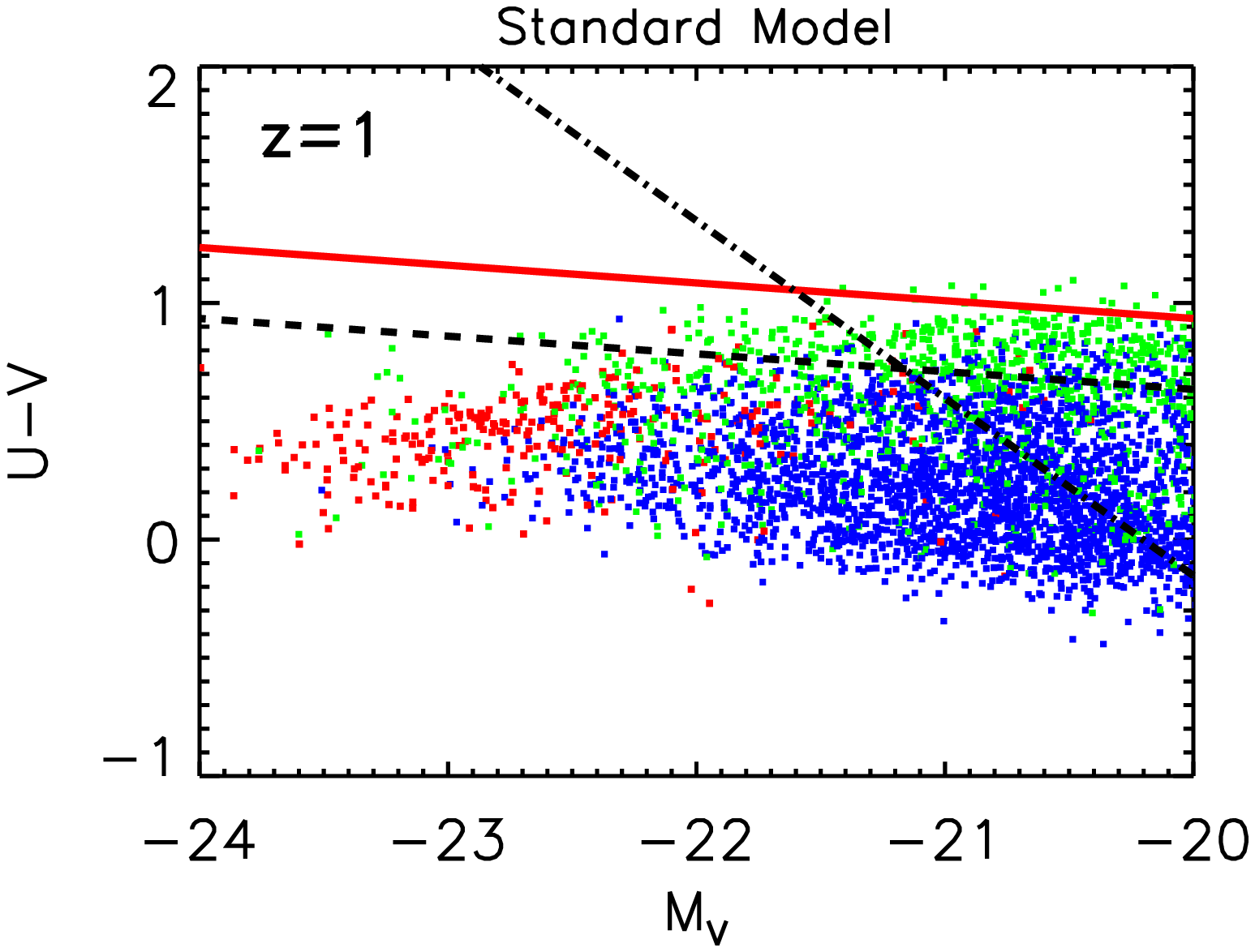,height=5.9cm,angle=0}
  }}
\end{minipage}\    \
\begin{minipage}{8.6cm}
\centerline{\hbox{
\psfig{figure=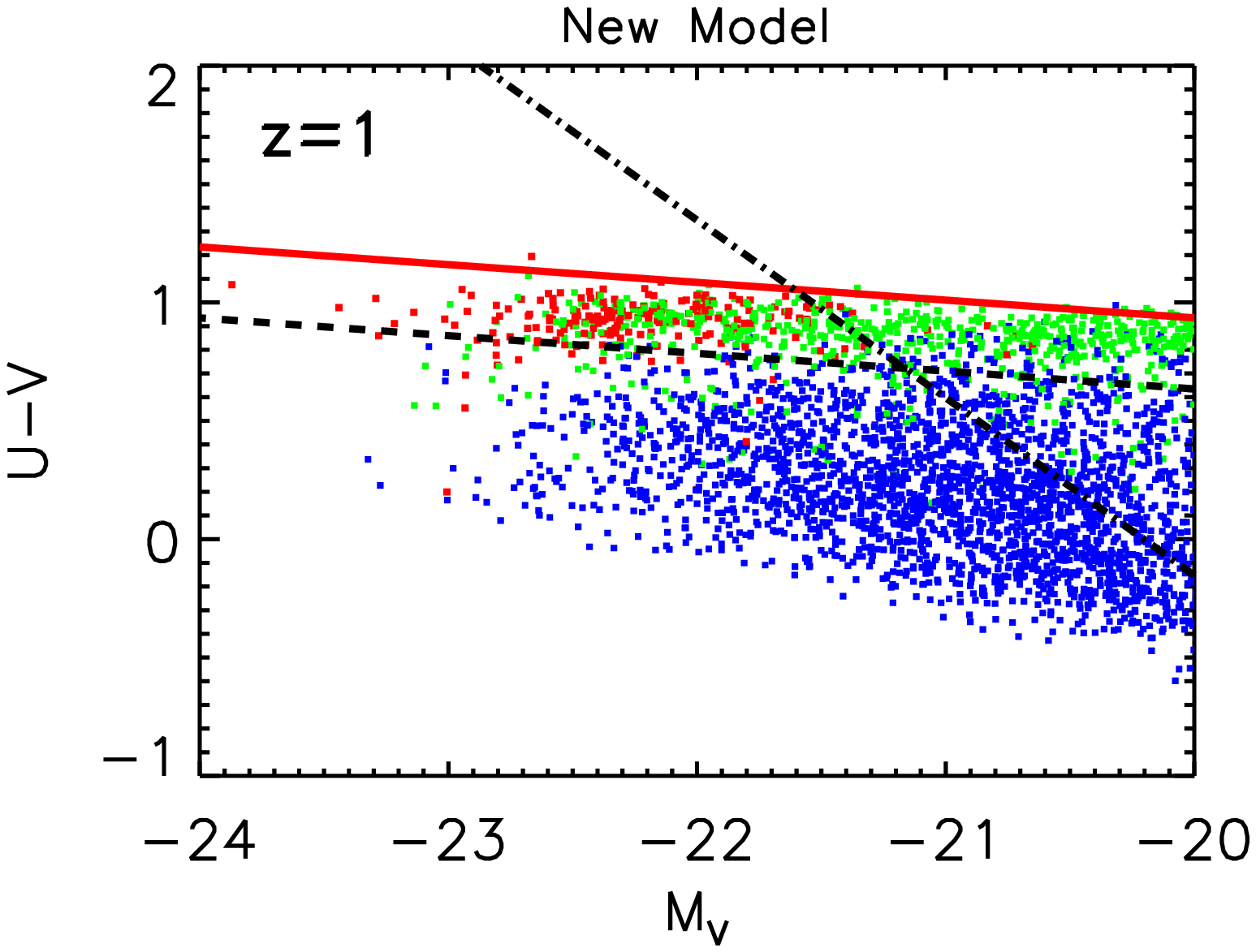,height=5.9cm,angle=0}
  }}
\end{minipage}\    \
\caption{Colour--magnitude diagram at $z=1$. $U-V$ versus $M_V$.
The ``standard" and ``new" models are the same as in \fig{z0_cm}.
The red solid line and the black dashed line refer to the
centre and the lower bound of the red sequence for the $1.0<z<1.1$ bin in the COMBO-17 data
 \citep{bell_etal04}.
The dot-dashed line marks the estimated completeness limit of that survey.}
\label{fig:z1_cm} 
\end{figure*}

\begin{figure*}
\noindent
\begin{minipage}{8.6cm}
  \centerline{\hbox{
      \psfig{figure=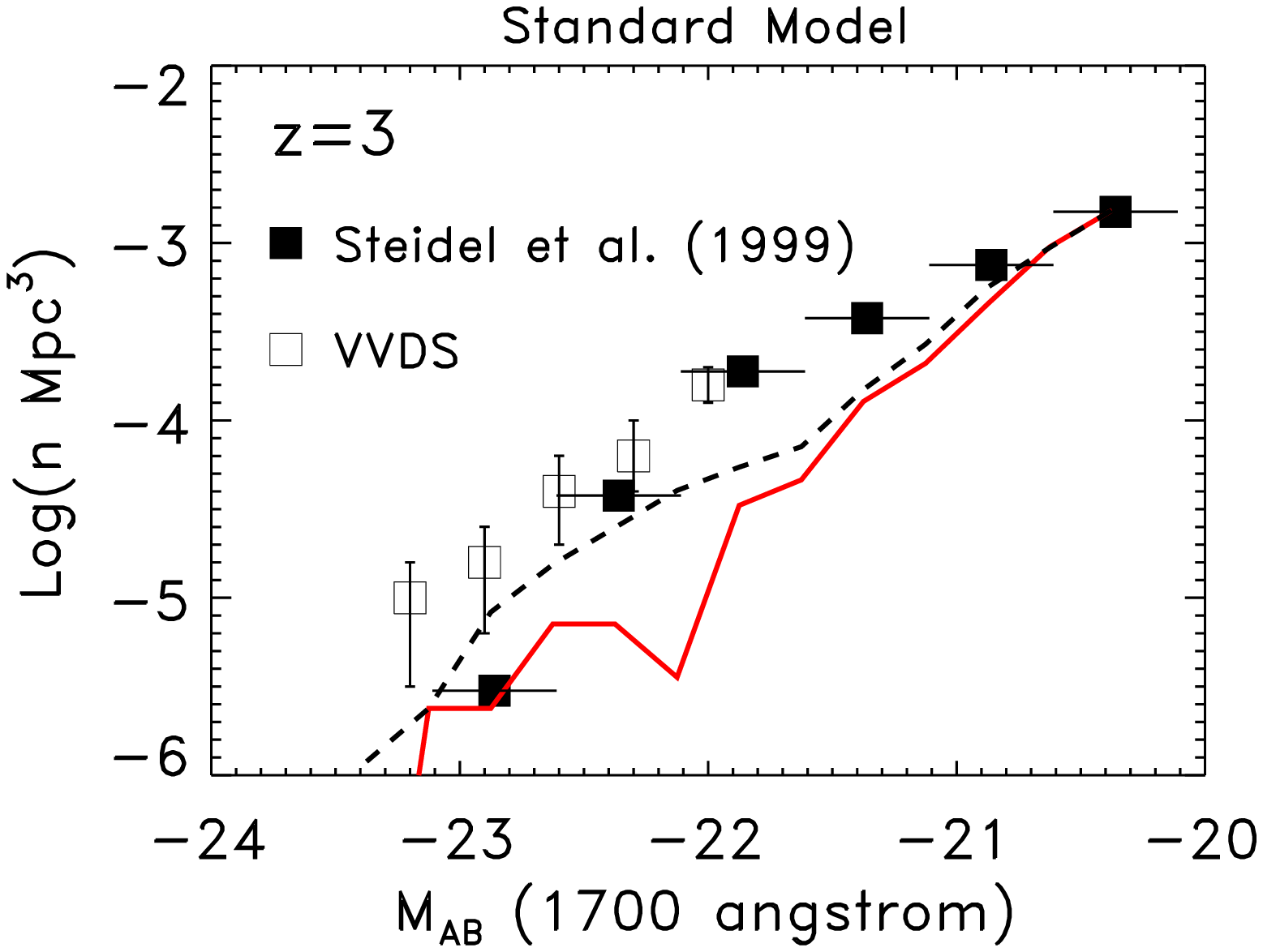,height=5.9cm,angle=0}
  }}
\end{minipage}\    \
\begin{minipage}{8.6cm}
\centerline{\hbox{
\psfig{figure=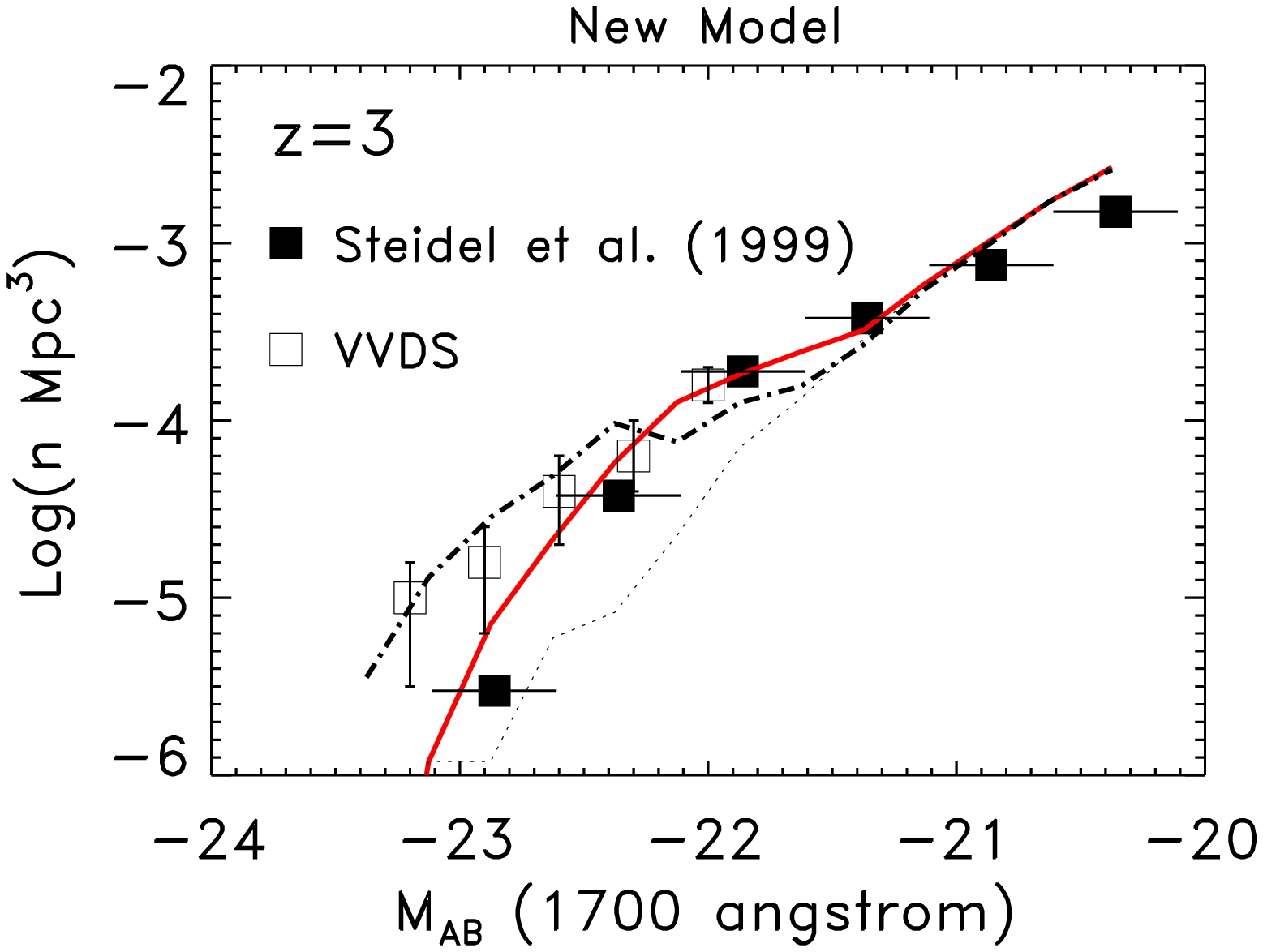,height=5.9cm,angle=0}
  }}
\end{minipage}\    \
\caption{Luminosity function at $z \simeq 3.1$.
The magnitude is rest-frame $0.17{\rm\,\mu m}$.
The model predictions (solid, red line), ``standard" (left) and ``new" (right),
compared to the data (symbols)
inferred from $R$-band observations of Lyman-break galaxies  
\citep{steidel_etal99}.
Superimposed are data from the VIMOS/VLT Deep Survey (VVDS).
The dashed line in the left panel
refers to an extreme variant of the ``standard" model
where all the gas cools instantly and feedback is practically shut off. Even with this extreme cooling
the model fails to form enough bright star-forming galaxies at $z\sim 3$,
demonstrating that a higher star formation efficiency is necessary at high $z$.
The dot-dashed line in the right panel refers to the ``new" model with
$z_{\rm c}=3.0$ rather than the fiducial $z_{\rm c}=3.2$, showing that
the high-$z$ predictions of the ``new" model are sensitive to the critical 
redshift after which $M_{\rm crit}=M_{\rm shock}$.
The thin dotted line in the right panel corresponds to 
$z_{\rm c}\rightarrow\infty$, namely $M_{\rm crit}=M_{\rm shock}$ for all $z$.
} 
\label{fig:z3_lf} 
\end{figure*}

\begin{figure*}
\noindent
\begin{minipage}{8.6cm}
  \centerline{\hbox{
      \psfig{figure=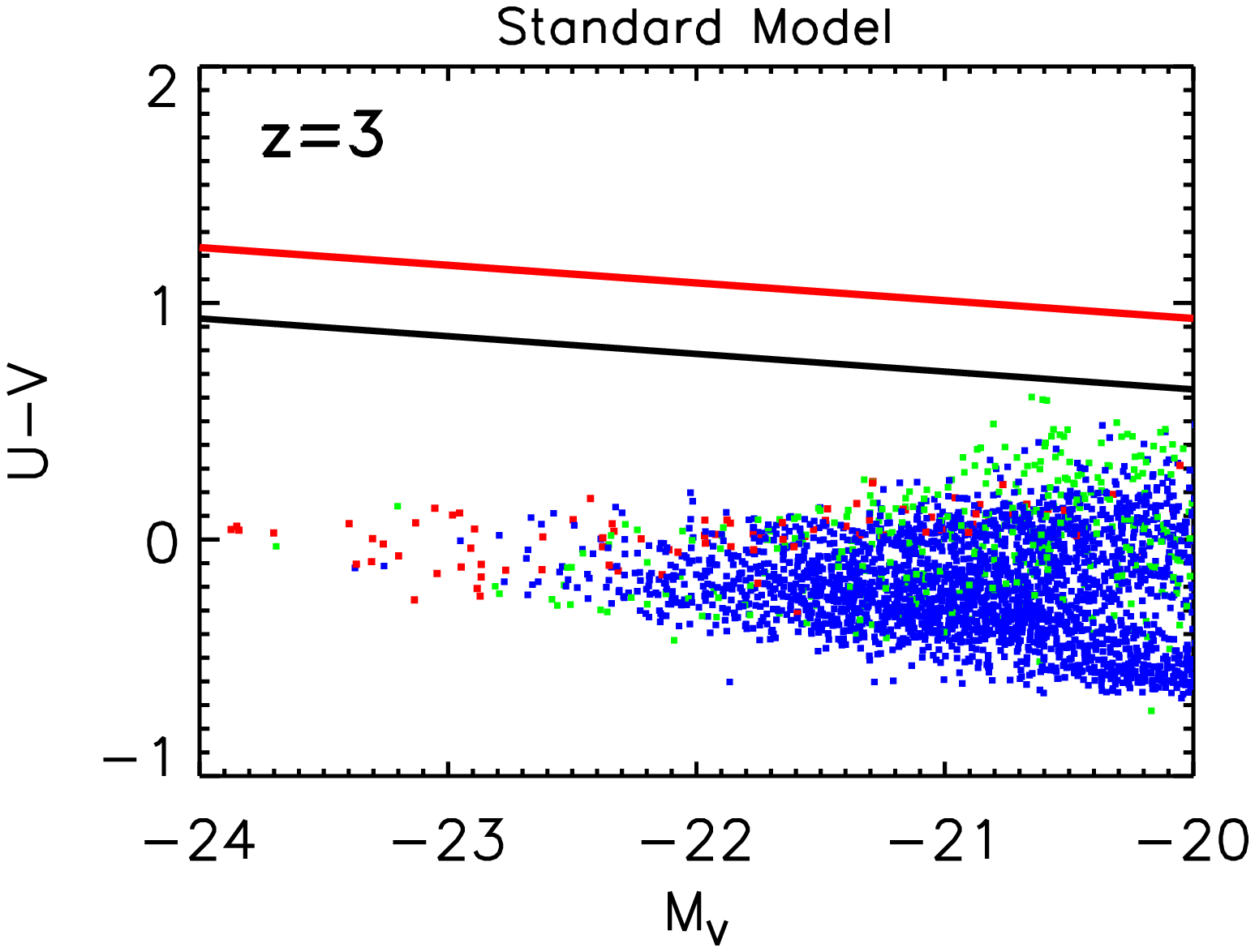,height=5.9cm,angle=0}
  }}
\end{minipage}\    \
\begin{minipage}{8.6cm}
\centerline{\hbox{
\psfig{figure=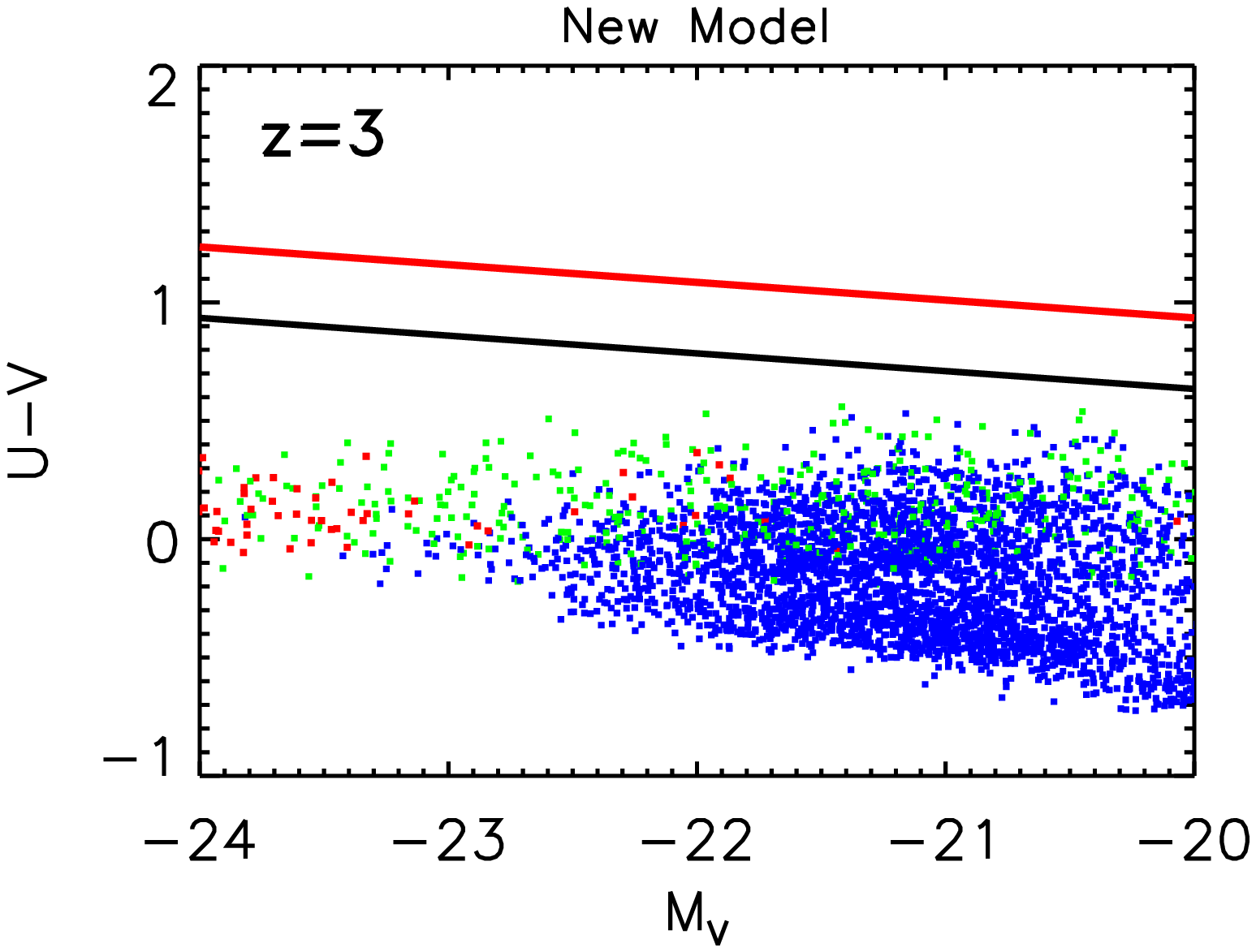,height=5.9cm,angle=0}
  }}
\end{minipage}\    \
\caption{Colour--magnitude diagram at $z=3$.
The ``standard" and ``new" models are the same as in \fig{z0_cm}.
The lines, shown for reference only, are the same as in \fig{z1_cm}  
describing the COMBO-17 red sequence at $z \simeq 1$.               
}
\label{fig:z3_cm} 
\end{figure*}

\begin{figure*}
\noindent
\begin{minipage}{8.6cm}
  \centerline{\hbox{
      \psfig{figure=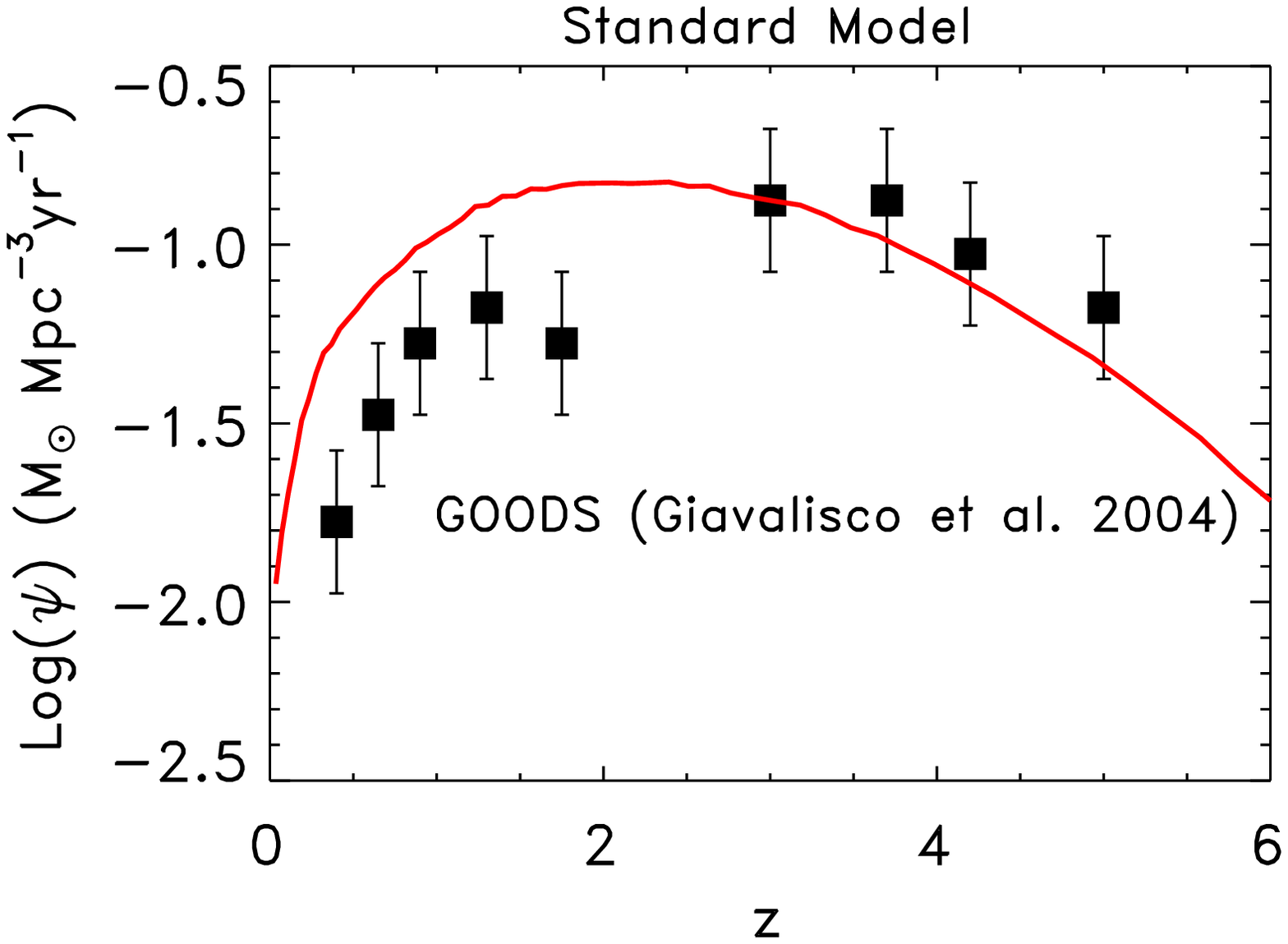,height=5.9cm,angle=0}
  }}
\end{minipage}\    \
\begin{minipage}{8.6cm}
\centerline{\hbox{
\psfig{figure=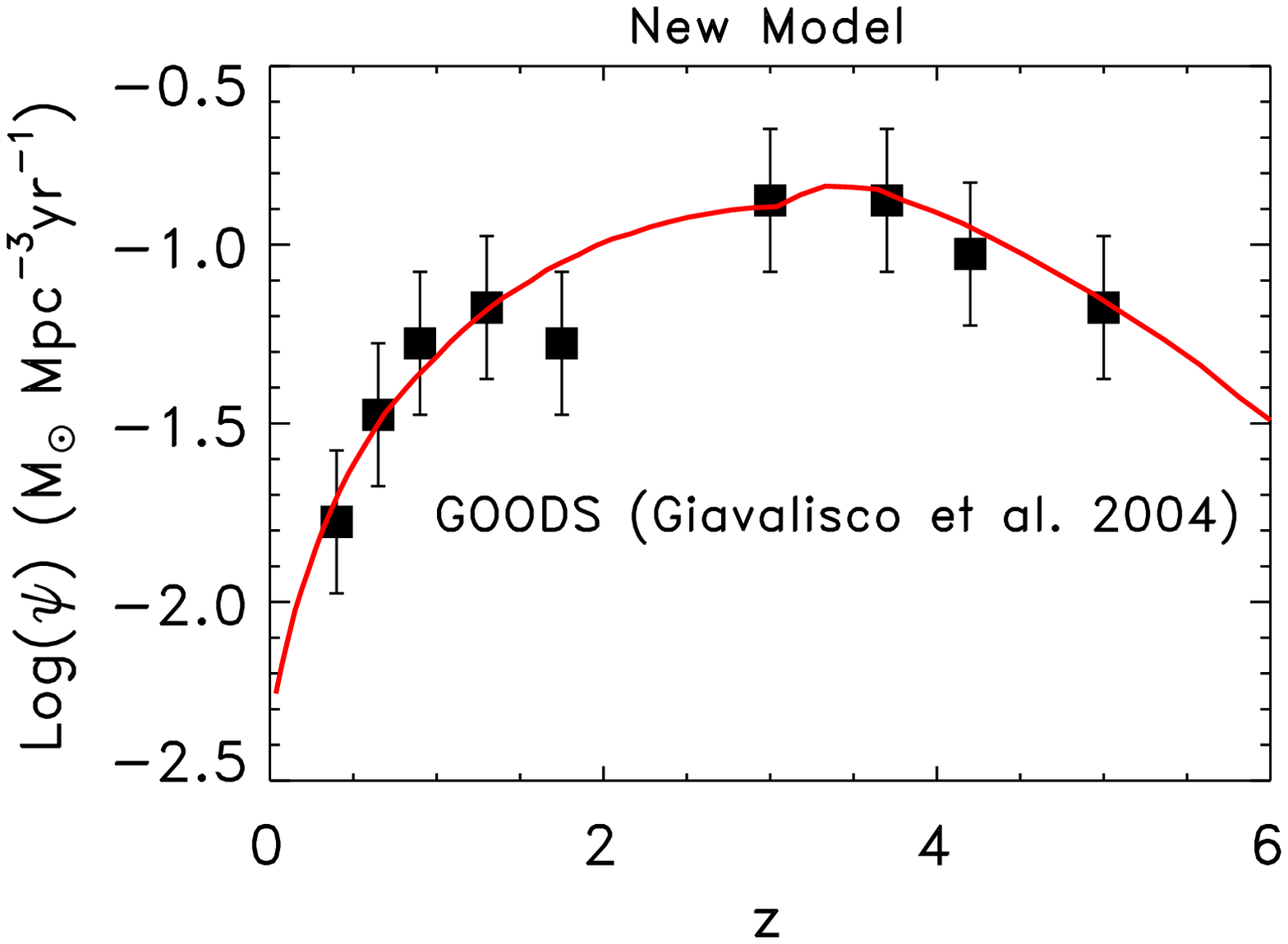,height=5.9cm,angle=0}
  }}
\end{minipage}\    \
\caption{Global star-formation history.
The model predictions (solid, red line), ``standard" (left) and ``new" (right),
compared to the data (symbols) from the GOODS survey \citep{giavalisco_etal04}.
While the ``standard" model overpredicts the star-formation density at $z<2$
the ``new" model provides an excllent match to the data at all redshifts.}
\label{fig:madau} 
\end{figure*}

\subsection{Cold streams and star formation at high redshifts}

The standard scenario under-predicts the number of bright galaxies in the 
rest-frame $0.17{\rm\,\mu m}$ luminosity function at $z\sim3$ (\fig{z3_lf}).
This is due to a slow conversion of gas into stars, as it is not cured
even when all the gas is allowed to cool in free fall from rest  
and supernova feedback is practically shut off. 
The shutdown of cooling in the most massive haloes will only accentuate 
the problem.
The problem can be cured if the star formation recipes derived from local
observations are revised to take into account the different way by which
galaxies accrete their gas at high $z$.
As discussed in \se{cold}, the cold gas is supplied by streams at the
virial velocity along the dark matter filaments.
The collision of the cold streams among themselves and with the
galactic disc produces bursts of star formation analogous to the
bursts resulting from the collision of two disc galaxies or cold gas clouds.
Under the conditions that allow a cold flow, these collisions are expected to 
produce isothermal shocks, and the
rapid cooling behind the shocks generates dense cold slabs 
in which the Jeans mass becomes small and stars can form efficiently.

The detailed physics of star formation under these conditions is yet to be 
worked out,
but this argument strongly suggests that the conversion of gas into stars
must have been more
efficient at high $z$ than it is now.
We mimic this effect in the ``new" model by introducing a 
$\sim(1+z)^{\alpha_*}$ increase in
the star formation efficiency (\equnp{sfr}), where $\alpha_*$ is a free 
parameter determined by fitting the luminosity function of Lyman-break 
galaxies at $z\sim 3$.

\section{Results}
\label{sec:rresults}

We have revised the GalICS SAM to accommodate the new features discussed above,
in particular
(a) the high star formation rate due to cold streams at high redshift,
(b) the abrupt shutdown of galaxy accretion and star formation due to 
shock heating in haloes above a critical mass,
and (c) the maintenance of this shutdown by AGN feedback that couples to the
hot gas.
This involves three new free SAM parameters, 
which are fine-tuned to reproduce the observational constraints.
They are:
(a) the shock-heating mass scale, $M_{\rm shock}$, 
(b) the critical redshift $z_{\rm c}$ prior to which $M_{\rm crit}$ is 
growing with $z$ (\equnp{mcrit}), and
(c) the power index $\alpha_*$ in the redshift dependence of the 
star formation rate (\equnp{sfr}).
The shock-heating scale is the single most important parameter 
determining the fit at low
and intermediate redshifts. As seen in the $M_r$--$M_{\rm halo}$ diagram of
\fig{z0_lum_mass}, the critical halos mass translates into a maximum 
luminosity for a star-forming galaxy, where the main inclined stripe  
marking the brightest galaxies intersects the vertical red line. 
This parameter is tuned to match the number density 
of blue ($u-r\lsim 2$) galaxies with $M_r\sim -22.25$ as observed in the SDSS. 
The other two parameters are determined by fitting the luminosity function 
of Lyman-break galaxies at $z\sim 3$ \citep{steidel_etal99}.
The obtained best fit values are
$M_{\rm shock}=2\times 10^{12}M_\odot$, $z_{\rm c}=3.2$, $\alpha_*=0.6$.

\begin{figure*}
\noindent
\begin{minipage}{8.6cm}
  \centerline{\hbox{
      \psfig{figure=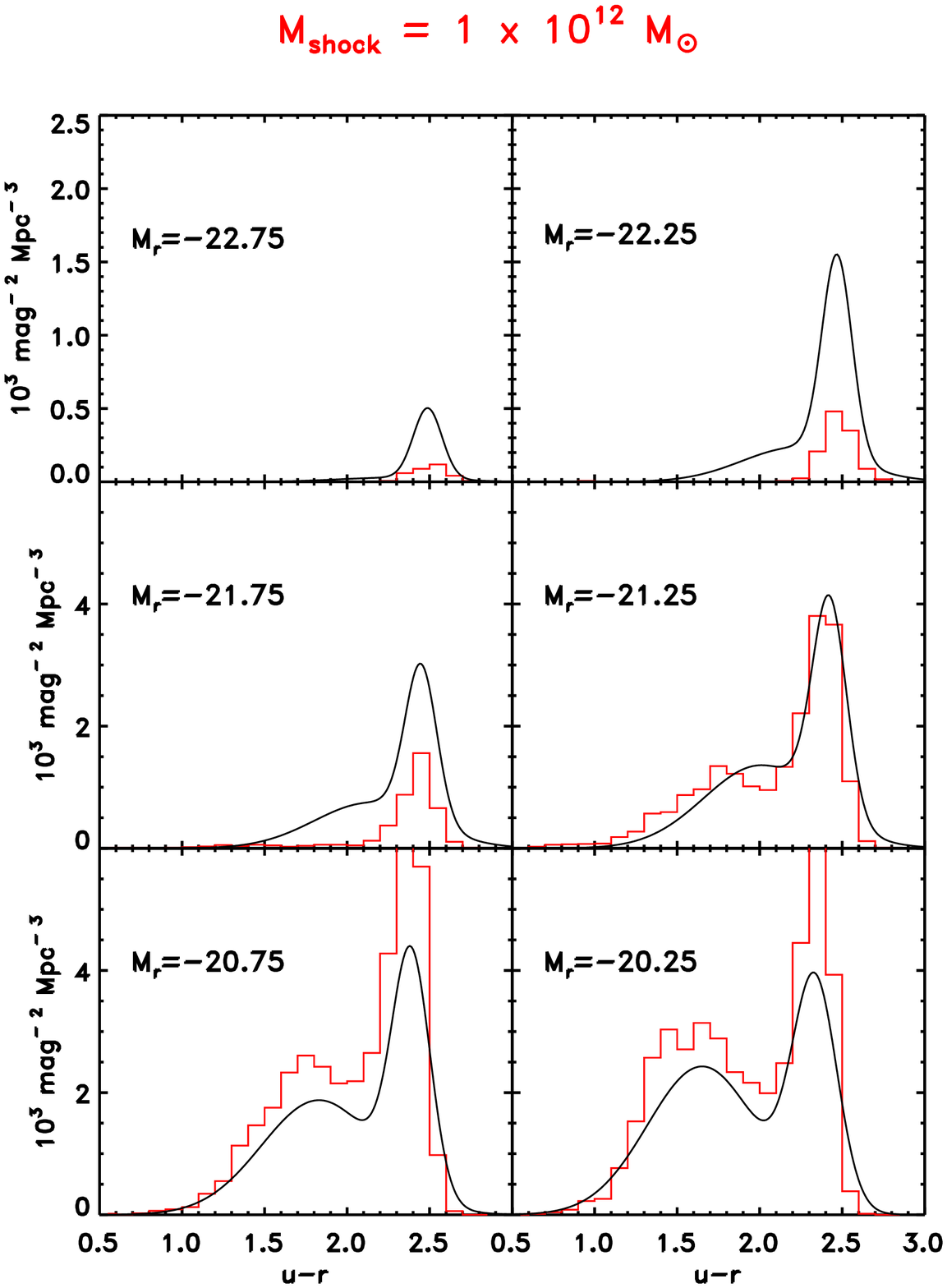,height=10.5cm,angle=0}
  }}
\end{minipage}\    \
\begin{minipage}{8.6cm}
\centerline{\hbox{
\psfig{figure=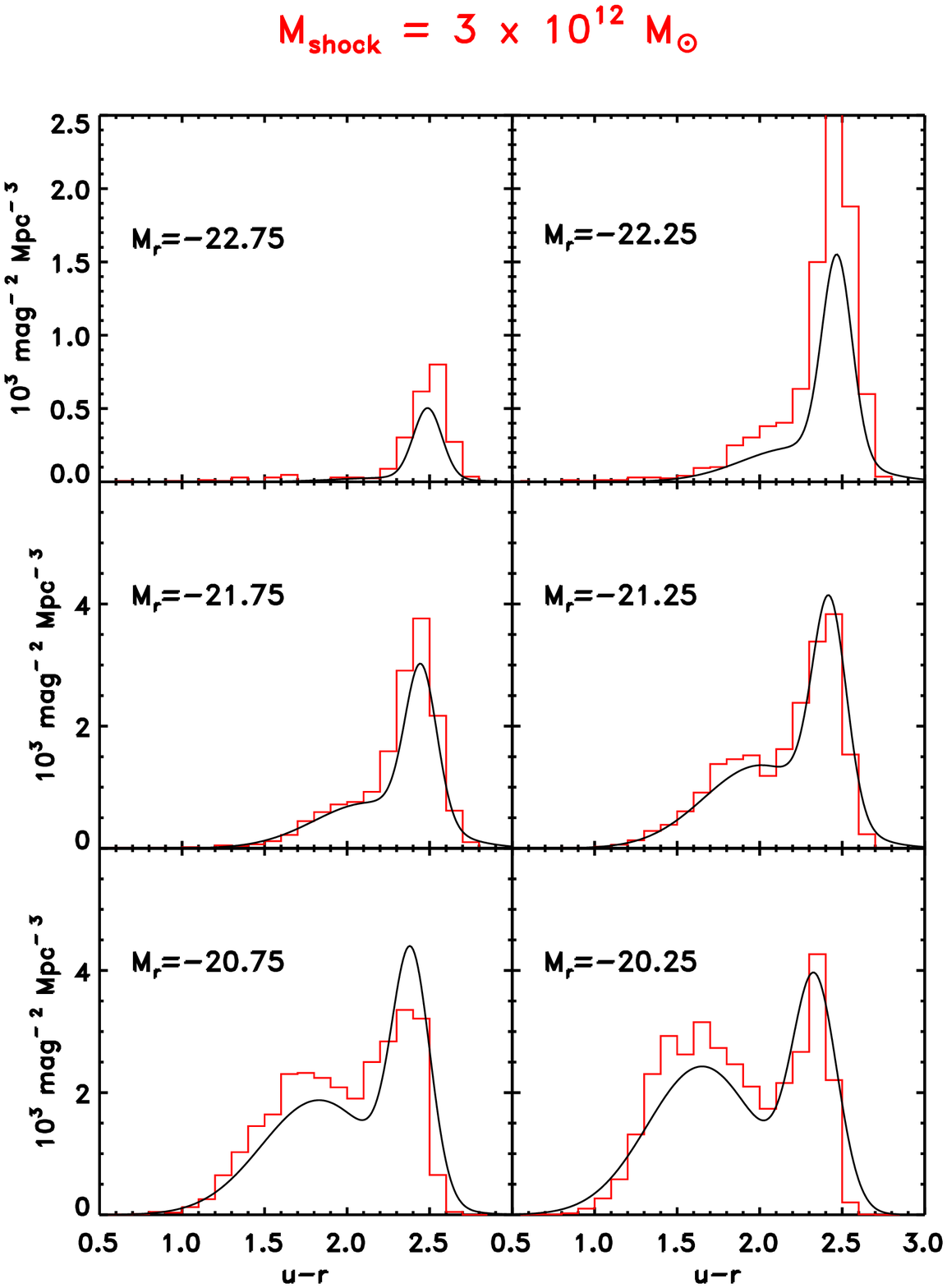,height=10.5cm,angle=0}
  }}
\end{minipage}\    \
\vskip 0.5cm
\noindent
\begin{minipage}{8.6cm}
  \centerline{\hbox{
      \psfig{figure=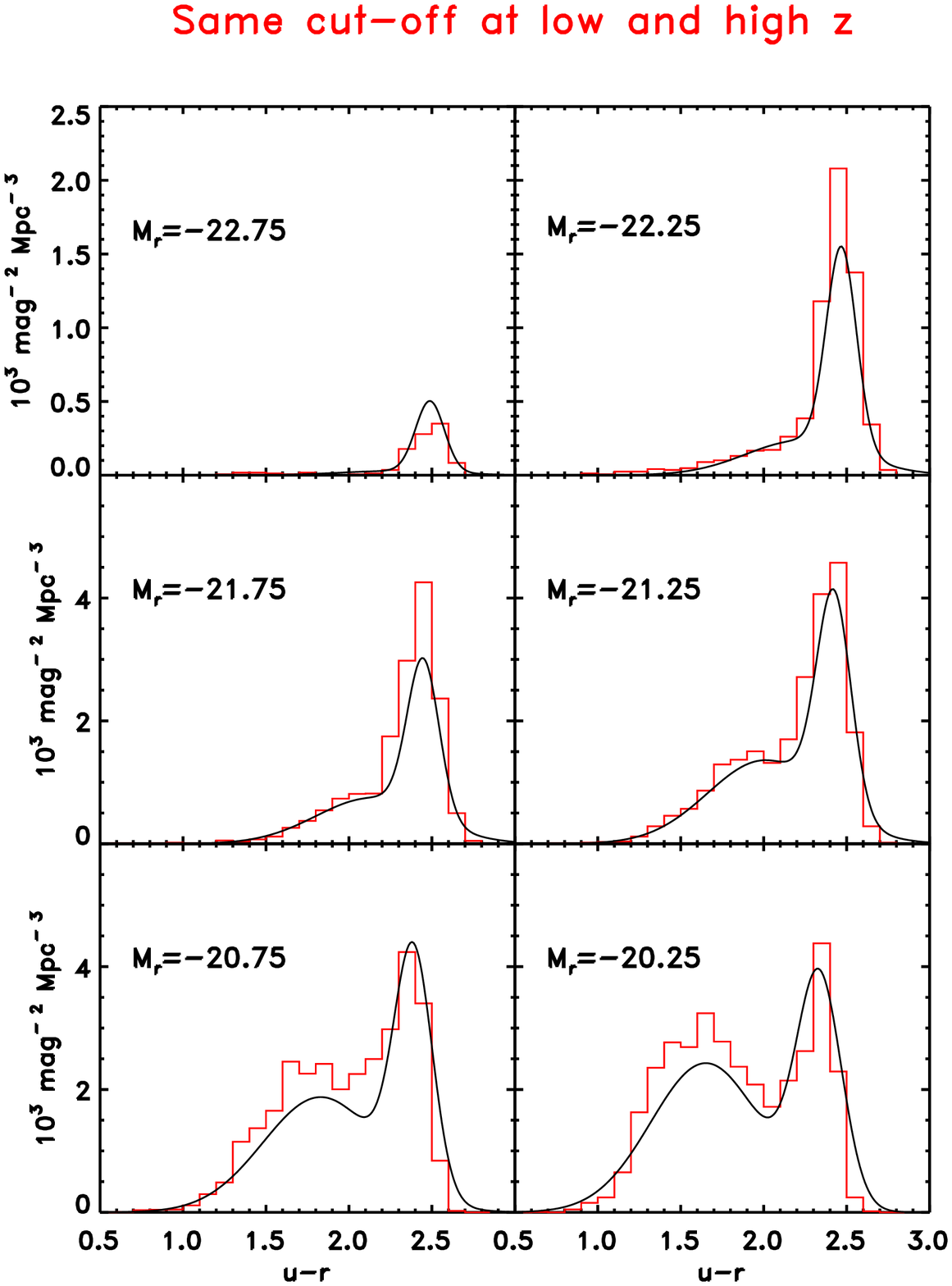,height=10.5cm,angle=0}
  }}
\end{minipage}\    \
\begin{minipage}{8.6cm}
\centerline{\hbox{
\psfig{figure=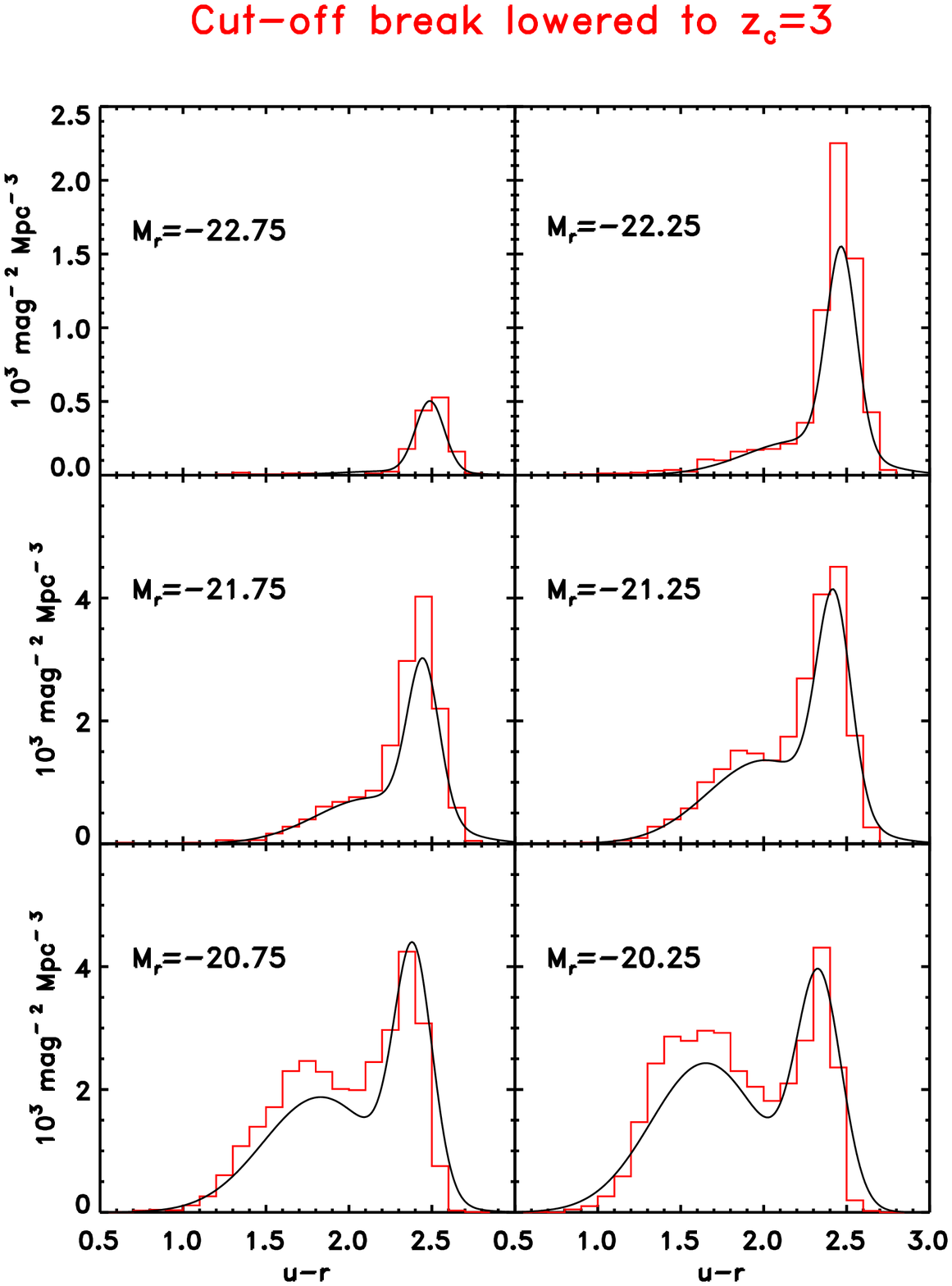,height=10.5cm,angle=0}
  }}
\end{minipage}\    \
\end{figure*}
\begin{figure*}
\noindent
\begin{minipage}{8.6cm}
  \centerline{\hbox{
      \psfig{figure=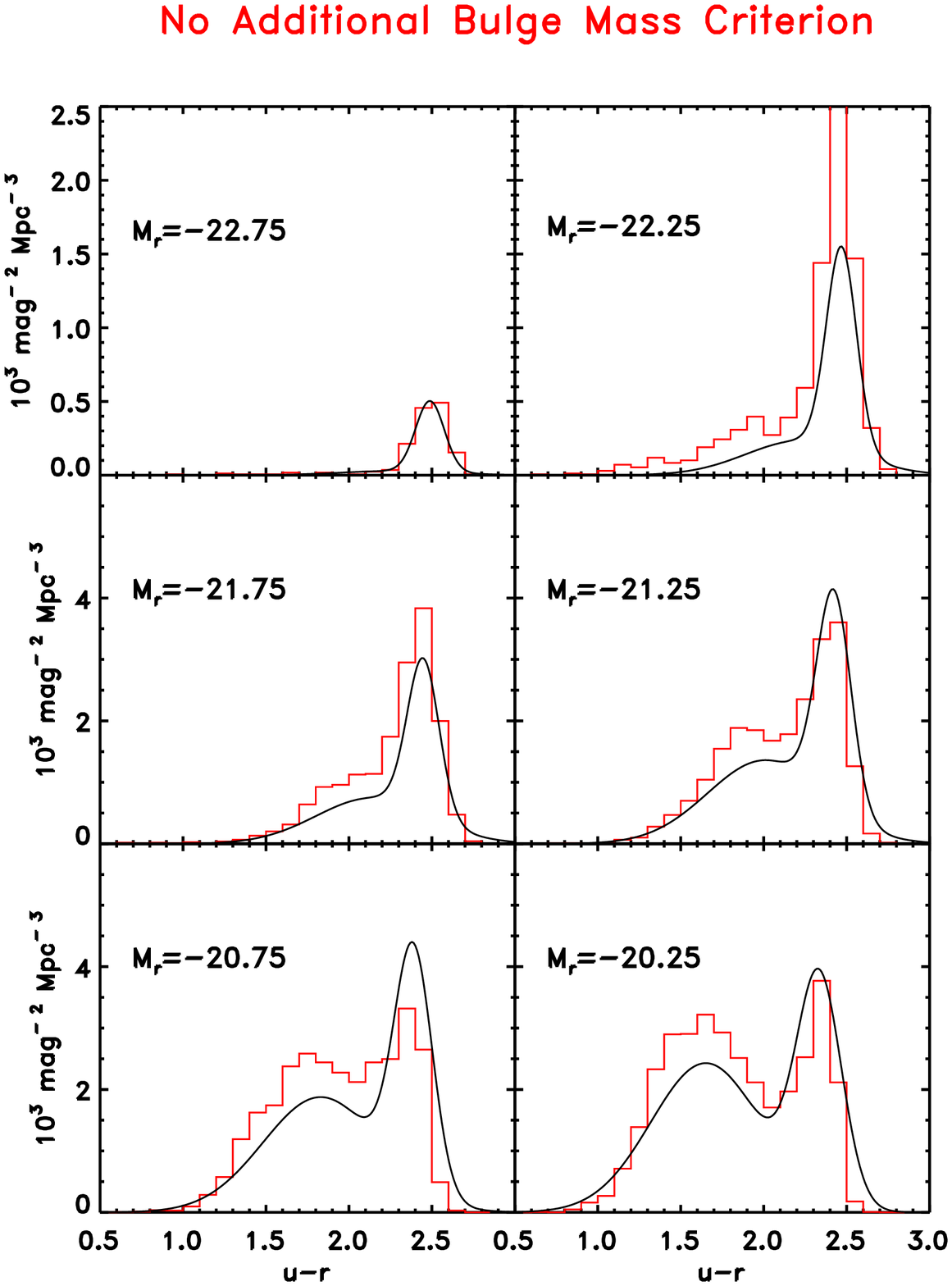,height=10.5cm,angle=0}
  }}
\end{minipage}\    \
\begin{minipage}{8.6cm}
\centerline{\hbox{
\psfig{figure=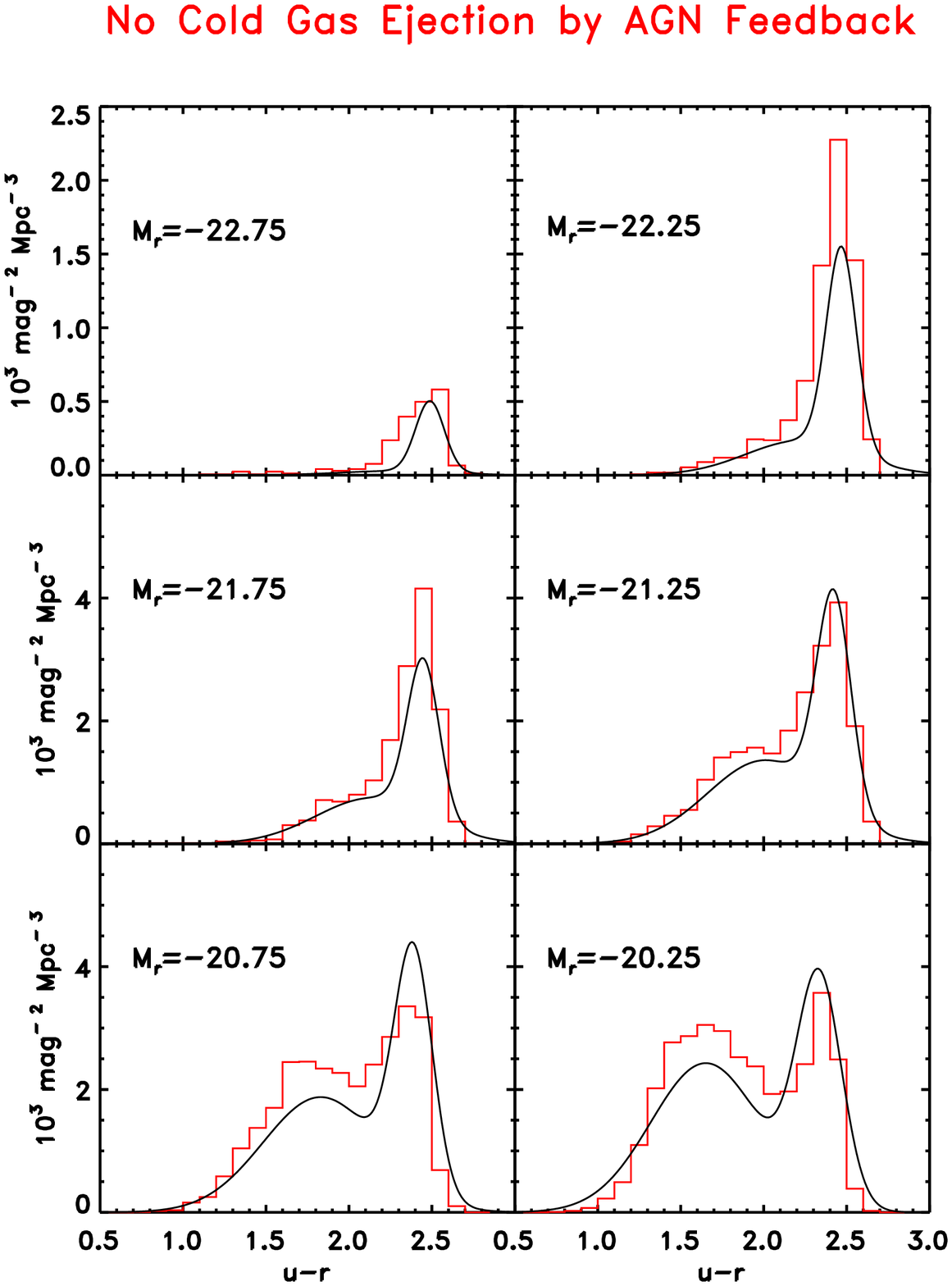,height=10.5cm,angle=0}
  }}
\end{minipage}\    \
\vskip 0.5cm
\noindent
\begin{minipage}{8.6cm}
  \centerline{\hbox{
      \psfig{figure=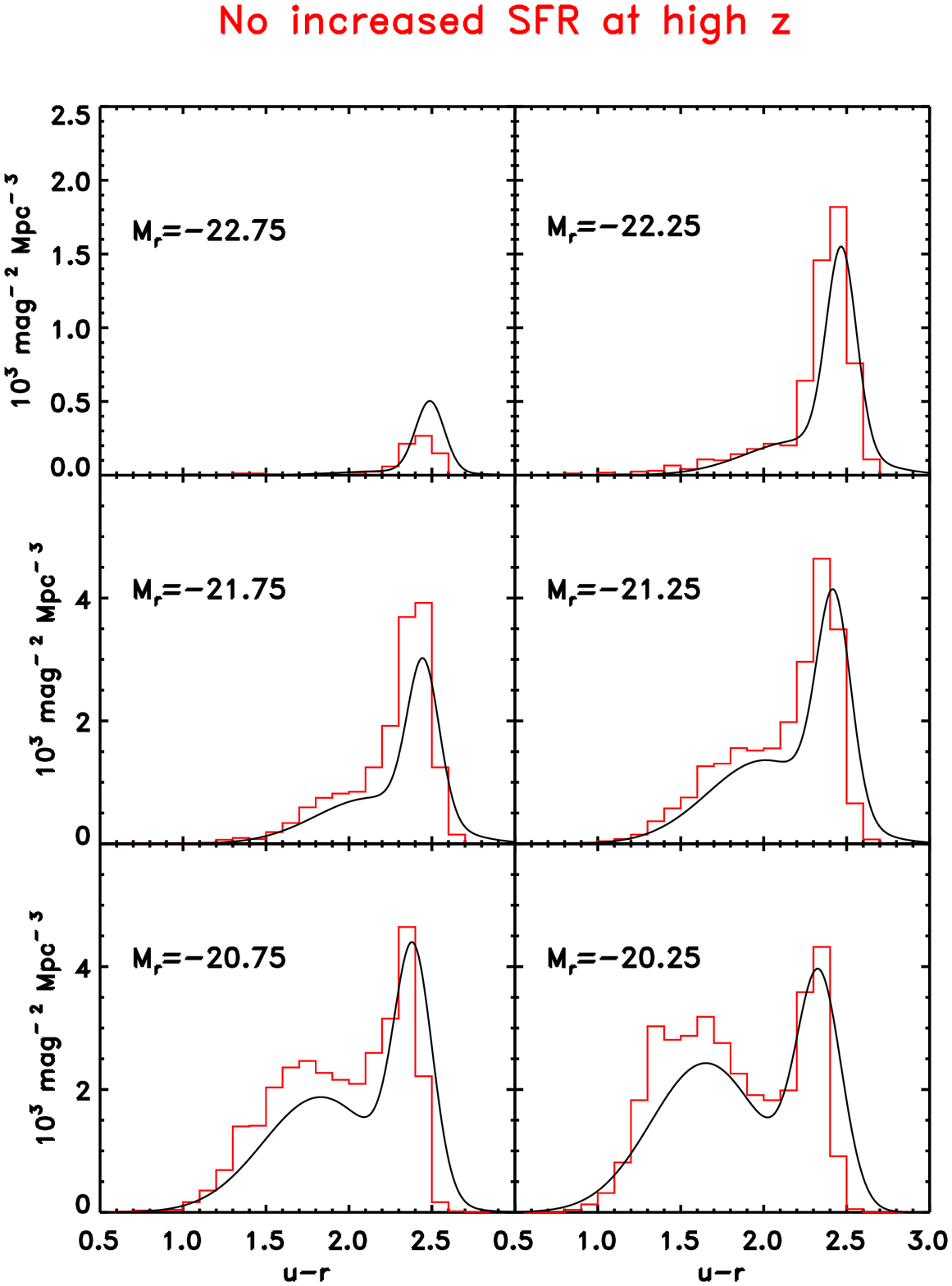,height=10.5cm,angle=0}
  }}
\end{minipage}\    \
\begin{minipage}{8.6cm}
\centerline{\hbox{
\psfig{figure=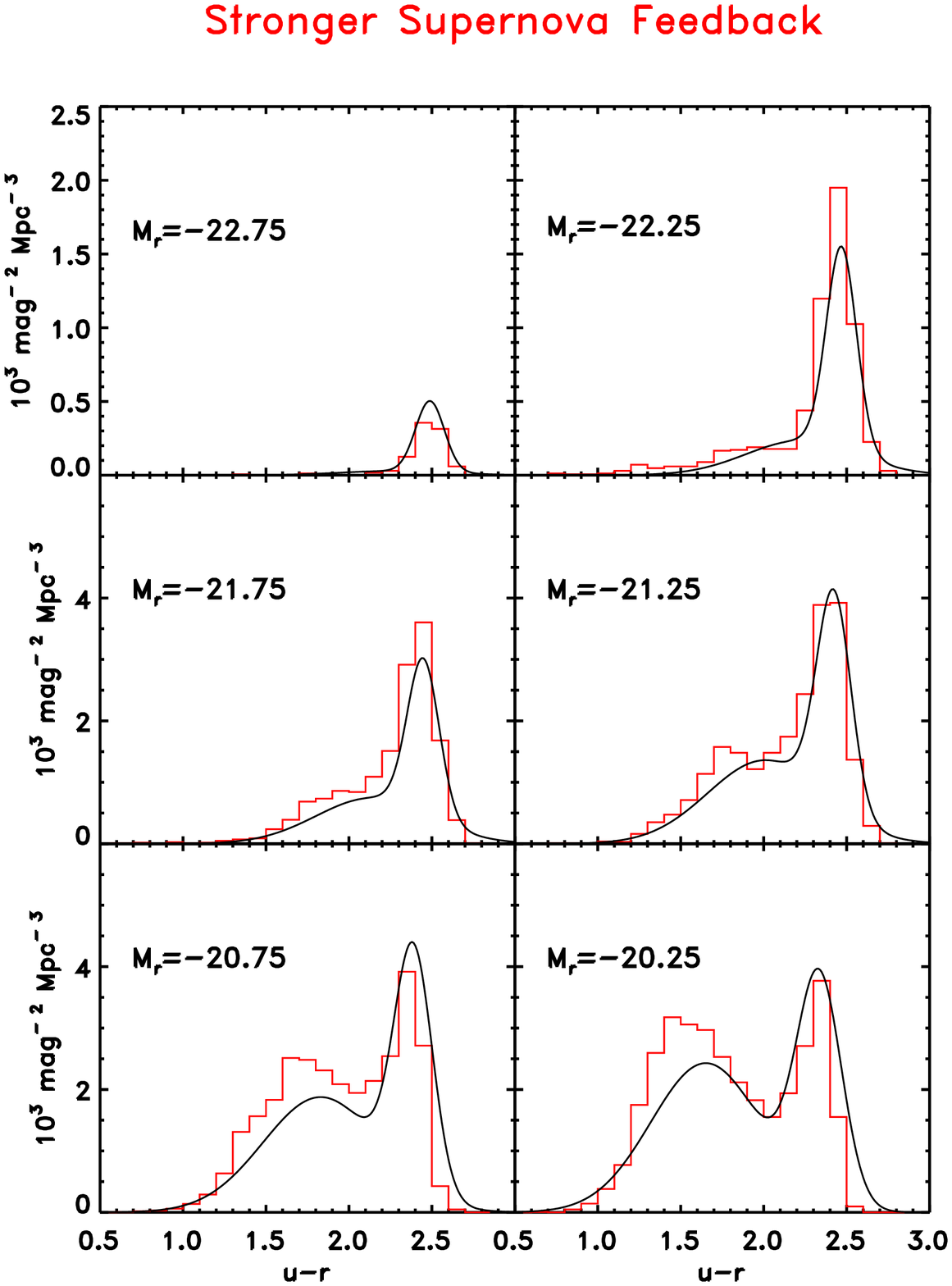,height=10.5cm,angle=0}
  }}
\end{minipage}\    \
\vskip 0.3cm
\caption{The effect of varying different model ingredients on the 
joint distribution of colour and magnitude at $z=0$.
Shown is the $u-r$ colour distribution in $r$-band magnitude bins for the
model predictions (red histogram) compared to the SDSS data 
(black smoothed histogram) from \citet{baldry_etal04}.
The models are the fiducial ``new" model of \fig{z0_hist} 
with one ingredient varied as follows:
(a) $M_{\rm shock}=10^{12\msun}$ (compared to $2\times 10^{12\msun}$ in the
fiducial model).
(b) $M_{\rm shock}=3\times 10^{12}\msun$.
(c) $z_{\rm c}\rightarrow\infty$, i.e., same $M_{\rm crit}$ at all $z$. 
(d) $z_{\rm c}=3$, somewhat lower than the fiducial value.
(e) No dominant-bulge requirement for shutdown of cooling and star formation.
(f) No ejection of the remaining cold gas when cooling is shut off.
(g) No increase in star formation efficiency at high redshift ($\alpha_*=0)$.
(h) An increased efficiency of supernova feedback ($\epsilon_{\rm SN}=0.25$).
}
\label{fig:robust} 
\end{figure*}

With this choice of the parameter values, we recover an excellent match to 
the $z=0$ luminosity function (\fig{z0_lf}) and to the bimodal 
colour-magnitue distribution in the SDSS (\fig{z0_cm} \& \fig{z0_hist}).
\Fig{z0_cm} shows that the blue sequence is properly truncated 
at $M_r\sim -22.25$.  The central galaxies of massive haloes 
are also positioned correctly in the SDSS red sequence colour range.
The colour histograms (\fig{z0_hist}) show that the
simulated joint colour--magnitude distribution agrees with the observed 
distribution both in shape and in normalisation.
We now reproduce the red population observed at $z\sim 1$ in the COMBO-17 
survey and the fact that the brightest galaxies are indeed on the red sequence 
(\fig{z1_cm}).
Finally, 
this model provides an excllent fit to the comoving number density of bright 
star-forming galaxies at $z\sim 3$ (\fig{z3_lf})
and to the observed cosmological star formation history (\fig{madau}).

\section{Sensitivity to model ingredients}
\label{sec:rrobust}

Here we explore the contribution of different ingredients to the 
success of the model in fitting the observed colour--magnitude distribution.

\subsection{The critical mass at low $z$}

The shutdown above the critical halo mass is the main new feature of our model,
and thus $M_{\rm shock}$ is the key parameter.
\Fig{z0_lum_mass} 
compares the distribution of galaxy luminosity versus
halo mass in the ``standard" model and our ``new" model.
Preventing cooling and star formation in haloes above $M_{\rm shock}$
keeps the magnitudes of these galaxies fainter than $M_r \sim -23$ 
The abrupt change in the mass-to-light ratio at the critical scale 
separates the red sequence from the blue sequence and sets an 
upper limit to the luminosity of galaxies in the blue sequence.
The best fit to the data, with $M_{\rm shock}=2\times 10^{12}M_\odot$,
is shown in \fig{z0_hist}. \Fig{robust} shows how by lowering (or raising)
$M_{\rm shock}$ to $10^{12}M_\odot$ (or $3\times 10^{12}M_\odot$) 
the model reproduces too few (or too many) galaxies at $M_r=-22.25$.

\subsection{The critical mass at high $z$}

At high $z$,
our fiducial ``new" model with $z_{\rm c} \sim 3$ allows star formation 
even in haloes above
$M_{\rm shock}$ (\equnp{mcrit}), mimicking cold streams in hot haloes. 
This change is responsible for the appearance in \fig{z3_cm}
of a population of bright galaxies with a high star formation rate at $z\sim 3$ 
(compare to \citealp{sawicki_thompson05}, 
while the red sequence of passively evolving galaxies has not formed yet.

The model predictions for Lyman-break galaxies (LBGs) are sensitive
to the value of $z_{\rm c}$.
When the increase of $M_{\rm crit}$ at high redshifts is not taken
into account, the predicted luminosity function at $z\sim 3$ falls short
at the bright end compared to that of bright LBGs
($z_{\rm c}\rightarrow \infty$ in \fig{z3_lf}).
On the other hand, when $z_{\rm c}$ is taken to be smaller than the fiducial
value of $3.2$, the predicted bright end of the luminosity function 
overshoots that of LBGs as observed by \citet{steidel_etal99}.
If, however, the actual comoving density of bright LBGs is somewhat higher, 
as indicated by the VIMOS/VLT Deep Survey (VVDS) team (private communication), 
then $z_{\rm c}=3.0$ becomes our best fit value (\fig{z3_lf}).
With this value, the match to the $r$-band luminosity function at $z=0$ 
becomes even better (\fig{z0_lf}), but in general the effect of $z_{\rm c}$ 
on the results at low redshifts is negligible (\fig{robust} c and d).

\subsection{A bulge criterion for shutdown}

Our fiducial model allows no cooling or star formation in galaxies where the
bulge is the dominant component. In practice, given the resolution limit 
of the N-body simulation, the critical shutdown scale
is reduced, in this case, to the resolution scale of 
$\sim 2\times 10^{11}M_\odot$.
This scale is significantly lower than the best-fit value of $M_{\rm shock}$,
but is only slightly lower than the critical halo mass as predicted
by theory (BD03; K05; DB06)
and as observed \citep{kauffmann_etal03b}.
As explained in \se{cold}, our $M_{\rm shock}$ is closer to an upper 
limit than to a mean value for the physical shock-heating scale.

When the effect of the bulge fraction is eliminated from the criterion 
for shutdown (\fig{robust}e), the model shows a small excess of bright 
galaxies ($M_r\sim-22.25$), both blue and red.  This is because the more 
efficient cooling allows more massive star-forming galaxies, which in turn
speeds up the merger rate due to more efficient dynamical friction.
We learn that the bulge criterion has some effect, but it basically serves
as a secondary criterion that helps fine-tuning the match to the data.
The fit is fairly adequate without it.

\subsection{Cold gas ejection in shock-heated haloes}

In the fiducial model, once the halo mass grows above the critical mass, 
all the gas remaining in the central galaxy is heated to the virial 
temperature and all modes of star formation are shut off. 
The idea is that when most of the gas is hot, AGN outflows 
heat, destroy or blow away the cold clouds as well.
When the cold gas is not ejected from massive galaxies but rather allowed 
to form stars later, the model predicts a small excess of blue galaxies 
in the brightest bin (\fig{robust}f). 
This is similar to the effect of eliminating the bulge criterion, but
even less noticeable.

\subsection{Efficient star formation at high redshift}

We have shown that increasing the star formation efficiency at high $z$ 
is necessary to reproduce the luminosity function of LBGs.
When the star formation is kept constant with redshift, instead,
star formation is postponed to later epochs.
The model predicts a shortage of galaxies in the brightest bin
and the colours become slightly too blue, but these are weak effects
(\fig{robust}g).

\subsection{Supernova feedback}
 
Our fiducial model for supernova feedback is similar to the model used in
most semi-analytic models based on \citet{dekel_silk86}.
It is more efficient in less massive galaxies with shallower potential wells.
The naive expectation is that a stronger supernova feedback would
make the galaxies redder by removing gas and stopping star formation.
Instead, we find that supernova feedback makes galaxies slightly bluer by 
ejecting metals from the galaxies into the IGM (\fig{robust}h).
The metal-enriched gas is later accreted onto the hierarchically assembled
more massive galaxies, and it is kept there as supernova feedback 
becomes less efficient. Thus, supernova feedback ends up transferring metals 
from low mass galaxies into massive galaxies. 

\smallskip
We conclude that the main element responsible for the improved match
of the model to the colour--magnitude data at low redshifts is the abrupt
shutdown above a critical mass of $\sim 10^{12}\msun$. All the other changes
in the model recipes, including the bulge criterion, are of secondary 
importance. They serve for fine-tuning the model in order to achieve a 
nearly perfect match.

\section{Correlations with Other Properties}
\label{sec:disc}

\begin{figure*}
\noindent
\centerline{\hbox{
\psfig{figure=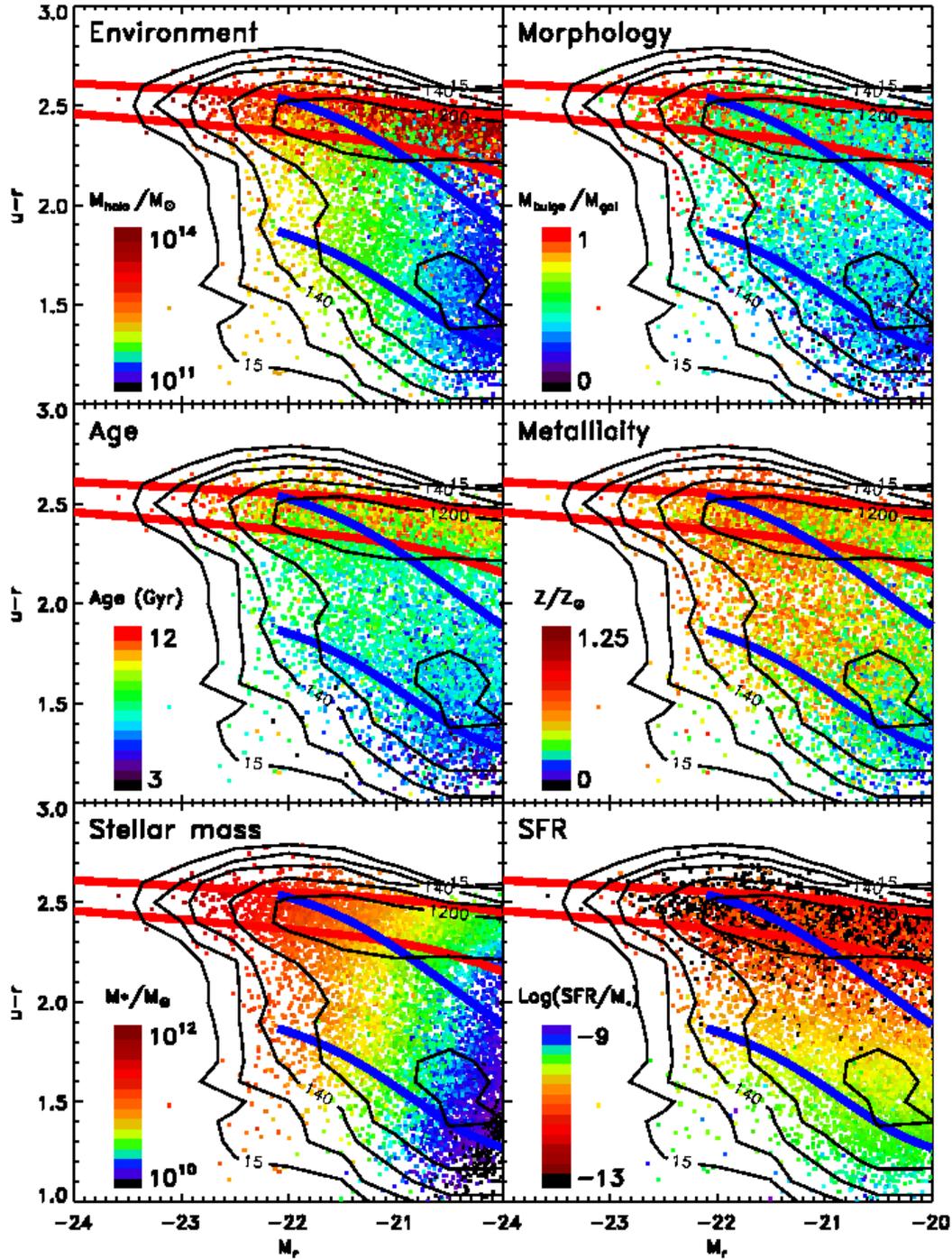,height=21.cm,angle=0}}}
\caption{Correlation of colour--magnitude with other properties in the ``new"
model at $z=0$.
As in \fig{z0_cm}, the points mark individual galaxies in the simulation
and the contours refer to their number density in the
colour--magnitude plane (${\rm mag}^{-2}{\rm Mpc}^{-3}$), 
while the blue and red lines mark the blue and red sequences in the SDSS. 
All the above are the same in all six panels.
The colour coding in each panel represents a different galaxy property as
follows:
(a) Mass of the host halo (log scale), which is correlated with the
    environment density.
(b) Morphology, quantified by the bulge-to-total mass ratio. 
(c) Mass-weighted stellar age.
(d) Mass-weighted stellar metallicity.
(e) Total stellar mass (log scale).
(f) Star formation per year per unit stellar mass (log scale).
Black corresponds to galaxies with no star formation at all.
}
\label{fig:environ}
\end{figure*}

The observed bimodality of the galaxy population, which is very apparent in the
colour--magnitude diagram, also involves all other global 
properties of galaxies. We already saw in \fig{z0_cm} that there is a strong
correlation between being blue or red and being below or above the critical
halo mass. In \fig{environ} we explore a variety of such correlations 
as predicted by the ``new" model at $z=0$.  In the same colour--magnitude 
diagram shown in each panel, colour refers to a different galaxy property.
The properties considered here include
environment density via halo mass,
bulge fraction, stellar age, stellar metallicity, total stellar mass and
star-formation rate.

A useful way to interpret the various correlations displayed in \fig{environ}
is to consider the evolutionary track of a typical galaxy in the 
colour--magnitude diagram.
Panels {\it a} and {\it e} show that the blue sequence is a sequence of 
growing halo mass and stellar mass from bottom-right to top-left.
As a halo is growing in mass, starting from the smallest mass resolved by 
the N-body simulation, gas is accumulating in the disc of the central galaxy 
and stars are forming.
The galaxy becomes brighter because its stellar mass increases, but it also 
gets redder because a growing fraction of the stellar mass is in old stars. 
In this evolutionary phase,
the galaxy moves along a stripe which roughly coincides with the blue sequence 
of the SDSS marked by the blue lines in \fig{environ}.

A galaxy reaches the top of the blue sequence when its host halo
mass becomes comparable to $M_{\rm crit}$. Soon after, it 
ceases to accrete gas, star formation shuts off, and the galaxy
evolves passively to become red and dead. 
Most galaxies never reach the top of the blue sequence because they merge 
into a halo much more massive than their own before their original halo 
exceeds the critical mass.  These galaxies are typically of low mass 
compared to the halo into which they have merged, and therefore their
dynamical friction time for sinking into the new halo centre is long.
They become the ``satellite" galaxies populating groups and clusters of 
galaxies.
Since satellite galaxies are assumed not to accrete gas, they fairly
quickly exhaust their remaining gas and become passive. 
Their brightness fades and their colour becomes redder, 
as the most massive, bluest stars are the first to die.
These evolutionary tracks can be deduced, for example,
from the stellar mass panel of \fig{environ}, where
galaxies evolve passively along the equal colour stripes, which
stretch roughly perpendicular to the blue sequence, from bottom-left
to top-right.
By mentally superimposing these diagonal equal-stellar-mass stripes 
on the panel of \fig{environ} referring to halo mass, one can verify that 
the evolution up along such  stripes involves an increase in halo 
mass, consistent with merging into a more massive halo.
A similar inspection of \fig{z0_cm} shows that this is indeed a transition
from a central galaxy to a satellite.
\begin{figure}
\centerline{\hbox{
\psfig{figure=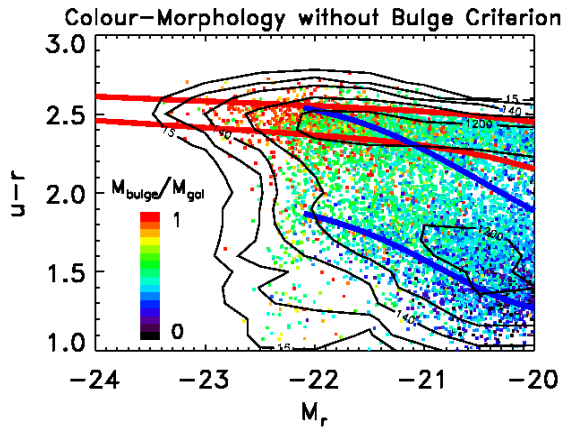,height=6cm,angle=0}}}
\caption{Morphology variations in a model with no bulge criterion
for cooling shutdown. The similarity to the bulge panel in \fig{environ}
indicates that the bulge criterion is of secondary importance to the
correlations between colour, luminosity and morphology.  
}
\label{fig:nobulge}
\end{figure}

Fading into the red sequence need not be the end point. 
Red galaxies, in hot massive haloes, do not accrete gas, but they can still
merge with other galaxies. 
The highest merger rates are for massive galaxies in the bright part of 
the red sequence,
as higher mass implies stronger dynamical friction. 
Thus the more massive a galaxy is when it leaves the blue sequence,
the more it will continue growing by merging after it has joined the 
red sequence.
Repeated mergers with other massive galaxies allow the formation of giant 
ellipticals with stellar masses of  $\sim 10^{12}M_\odot$,
which would be unattainable through simple accretion of gas in the 
presence of a limit on cooling.
Most of this growth is due to dissipationless merging within the red sequence.
This interpretation of the role of merging and passive evolution
is supported by the dominance of central galaxies in the bright part of 
the red sequence and of satellite galaxies in the faint side, separated
roughly at $L_*$, the characteristic luminosity of a galaxy at the top 
of the blue sequence (\fig{z0_cm}). 
It is also supported by the variation of morphology with position in
the colour--magnitude diagram (\fig{environ}b).
Blue galaxies are mainly spirals, with bulge-to-disc ratio that increases 
with luminosity. The morphological composition along the red sequence
also varies from mostly spiral galaxies at $M_r\sim -20$
to mostly elliptical galaxies at $M_r\sim -22.5$,
evidence for an increasing merger rate from faint to bright galaxies.

The correlation of colour and magnitude with morphology is a 
generic
outcome of the merger scenario and is not an artifact of the bulge 
criterion for shutdown. 
This is demonstrated in \fig{nobulge}, in which the bulge criterion 
was eliminated.  The central galaxies of group and cluster size haloes, 
shown as red 
dots at the bright-end of the red sequence 
in \fig{z0_cm} are those with the most  intense merger history 
and therefore with the highest bulge-to-disc ratio.
A qualitatively similar 
correlation with morphology is 
valid already in the ``standard" model without cooling shutdown. 

The presence of two populations within the red sequence manifests itself 
also in the environment panel of \fig{environ}.
The galaxies that live in the most massive haloes ($10^{13-14}\msun$) dominate
both the bright end and the faint end of the red sequence, but the middle
range between $M_r = -21$ and $-22.5$ shows many middle-of-the-range haloes of
$10^{12-13}\msun$ and even smaller.
This means that as the halo mass grows in time the galaxy can evolve according
to one of the following two tracks: it may either keep on growing until it 
becomes a bright E/S0 galaxy or fade to become an early-type satellite
with a signifcant disc.
This prediction of our model is consistent with the SDSS results
\citep{blanton_etal06},
where the number density in the $1h^{-1}\,$Mpc environment of galaxies
is measured to peak both at the bright end and at the faint end of the 
red sequence while the morphology shows a gradient along the red sequence.

We learn from the star formation rate panel of \fig{environ} that 
model galaxies are red because they are not forming stars and thus lack a 
young stellar population.
Indeed, the distinction between the red sequence and the blue sequence is 
predominantly due to age rather than metallicity, as can be deduced
by comparing the gradients at a given luminosity in the age panel and
the metallicity panel of \fig{environ}.
However, the colour gradients along the red sequence and along the blue
sequence themselves are driven by the metallicity gradient more than by the 
age gradient.


\section{Conclusion}
\label{sec:conc}

We have addressed the origin of the robust division of the galaxy 
population into two major types: the blue sequence of gas-rich
galaxies with a young stellar population and the red sequence of
gas-poor galaxies dominated by old stars. While the blue sequence is
truncated beyond a characteristic stellar mass of $\sim 3\times 10^{10}\msun$,
the red sequence extends to higher stellar masses.
This bimodality is strongly correlated with the morphological type
and with the galaxy density in the environment.
Large galaxy surveys have revealed that the bimodality is reflected
in almost every other property of galaxies and that the separation between 
the two types is fairly sharp.
High-redshift surveys show that a fraction of the red sequence is in place
already at $z \sim 1-2$, and that massive star-forming galaxies exist 
at $z\sim 2-4$.

We find that models of galaxy formation can 
successfully 
reproduce the bimodality
once a complete shutdown of cooling and star formation is imposed in 
haloes above a critical mass of $\sim 10^{12}\msun$ starting at $z \lsim 3$. 
The time evolution of the bimodality is reproduced when allowing
partial cooling and efficient star formation above the critical mass
at higher redshifts.
By incorporating these simple new features in hybrid semi-analytic/N-body
simulations, we have achieved an unprecedented simultaneous match to the joint 
distributions of galaxy properties at low and high redshifts, practically
confirming the qualitative predictions of Dekel \& Birnboim (2006).

The motivation for the sharp transition at a critical halo mass
comes from theoretical analysis (BD03; \citealp{binney04}; DB06)
and cosmological simulations (K05; DB06)
showing a sharp transition from galaxy build-up by cold filamentary
flows in small haloes to virial shock heating of the halo gas in 
massive haloes. 
The abrupt shutdown in the supply of cold gas for star formation,
due to the abrupt appearance of stable virial shocks
as opposed to the smooth variation of cooling time with mass, 
is responsible for the very red colours of the red sequence,
the colour gap between the red and the blue sequence,
and the sharp truncation of the blue sequence above $L_*$.
 
An important input for the success of the model is that, once the gas 
is heated to the virial temperature, it never cools. In this model
shock heating serves as the trigger for efficient AGN feedback, which
maintains the gas hot. 
The long-term effect may be associated with
a self-regulated accretion cycle maintained by the coupling between the 
central black hole and the hot gas.  The luminous quasar mode is 
associated with the rapid build-up of massive black holes probably due to the 
high gas inflow rate triggered by galaxy mergers.  The resulting high 
output power may be important in terminating star formation and ending 
the quasar phase itself \citep[e.g.,][]{dimatteo_etal05}, but this
phase is short-lived and cannot serve to maintain the gas hot for long periods. 
The self-regulation is achieved when the accretion rate drops to an 
equilibrium value at which the energy ejected from the 
black hole and absorbed by the gas compensates for the radiative losses 
(Cattaneo 2006).  The mechanical efficiency of this mode is close to 
100\%, while in the quasar mode most of the output power goes into radiation.
In the quiet mode, the dense clumps of cold gas are not swift in 
responding to the external injection of mechanical energy --- they can be 
destroyed or blown away only when the AGN is sufficiently powerful as 
in a quasar.  On the other hand, the hot gas, which feeds the black hole 
in a gradual fashion, is more dilute, and can therefore respond effectively 
to the injected energy.  Therefore, the self-regulated mode of hot
accretion may be responsible for a significant fraction of the mechanical
energy affecting the gas, even if the hot gas contributes only a small 
fraction of the final black hole mass.

In a parallel work, \citet{croton_etal06} have implemented a related 
semi-analytic model, where they focus on the growth 
of black holes and AGNs as feedback sources.  In particular, they assume 
a transition from rapid to slow cooling and associate the 
accretion of cold and hot gas with an `optical' and a `radio' mode of black-hole growth, respectively. 
The latter provides the relevant AGN feedback process and determines the gradual shutdown cooling
in massive haloes.
We focus instead on the thermal properties of the gas as the trigger for
an abrupt shutdown above a threshold halo mass, which translates into
the bimodality scale in the galaxy properties. The key to our successful 
modelling of the galaxy properties is this robust shutdown, regardless of 
the details of AGN feedback.  This simple addition to the model results in 
an unprecedented detailed match to the data at low and high redshift.

Our ``new" picture introduces a strong correlation between the spectral galaxy
type and the host halo mass, while the morphological type is driven by 
the galaxy merger history as in the ``standard" model. 
Thus, while the new model gas fraction and colour of massive galaxies 
are different, the morphological distribution and the correlation between 
morphology and environment density are reproduced as in the standard model. 
The natural correlation between the mass of the host halo and the density 
of the environment, according to the ``halo model" of galaxy distribution 
\citep{yan03_hod_2df,zehavi04_hod, krav04_hod},
is reflected in the correlation between spectral and morphological type. 

\begin{figure}
\noindent
\centerline{\hbox{
\psfig{figure=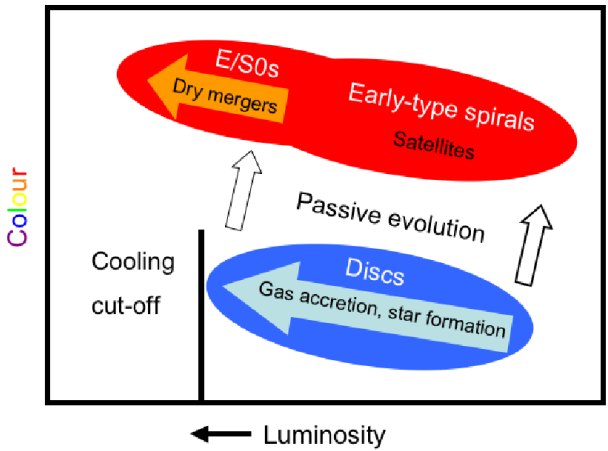,height=7.4cm,angle=0}}}
\caption{A schematic sketch of galaxy evolution. 
Galaxies grow along the blue sequence through cold filamentary flows 
until they stop accreting gas because their host halo has grown above 
the critical shock-heating mass.
At that point they move into the red sequence by becoming red 
and fading in luminosity. 
The shutdown of cooling above the critical halo mass sets an upper limit to
the luminosity of blue galaxies and explains the characteristic galaxy 
luminosity $L_*$.
Galaxies get to the E/S0 bright end of the red sequence either by
dissipationless (`dry') mergers along the red sequence,
or by gas-rich (`wet') mergers from the top of the blue sequence. 
This simple cartoon ignores the complications associated with the 
temporary reddening by dust extinction in  
star-forming galaxies.
}
\label{fig:scheme}
\end{figure}

A few comments on the evolution in the colour--magnitude diagram,
as summarised in \fig{scheme} (see also Dekel \& Birnboim 2006).
Spiral galaxies grow discs by cold filamentary streams at the centres of 
haloes below the critical mass.
As the accreted gas is converted into stars, the galaxies become 
brighter and redder along the blue sequence.
A galaxy leaves the blue sequence either when its halo grows above the 
critical mass or when it merges into such a halo and becomes a satellite 
galaxy.
In both cases, the galaxy ceases to accrete gas from its hot environment 
and rapidly turns red and dead.
Small satellite galaxies tend to fade as they redden and turn into
the red discs observed at the faint end of the red sequence 
\citep{blanton_etal06}.
Big central spirals whose haloes have grown above the critical mass
continue their growth along the red sequence through dissipationless 
mergers and evolve into the giant ellipticals at the bright end of the 
red sequence.
Gas-rich mergers may also exhaust the cold gas reservoir of the merging 
galaxies and transform two blue spirals into a red elliptical,
but this is a less frequent path in the formation of massive 
early-type galaxies.
This picture implies that the colour difference between the blue 
and red sequences is primarily due to stellar age, while  
the colour--magnitude gradient within the red sequence is
largely a metallicity effect, since deeper potential wells 
retain their metals more effectively.
The colour--magnitude gradient along the blue sequence is a combination 
of the two effects.

A detailed study of the origin of the red sequence via evolutionary
tracks in the colour--magnitude diagram is described by  
Cattaneo, Dekel \& Faber (in preparation).
At high redshifts, the enhanced star-formation rate that we have
incorporated in galaxies of all masses helps curing the difficulty of 
the standard scenario in forming enough stars in massive galaxies at 
$z\sim 2 - 4$, as indicated by the observations of LBGs and SCUBA sources.
This new ingredient is motivated by the build-up of galaxies via nearly 
supersonic filamentary flows (BD03; K05; DB06). 
The collisions of these streams among themselves and with the
central discs are likely to trigger efficient bursts of star formation,
in analogy to collisions of discs or cold clouds.
However, the fate of cold streams in hot haloes and the resulting 
star formation are yet to be studied in detail.

We note that our scenario helps explaining the ``downsizing" of the
formation of massive galaxies, where more massive galaxies tend to form 
stars earlier and over shorter periods.
The most massive galaxies, those above the shock-heating mass, have formed  
their stars at $z>2-3$ by cold flows in hot media, and
stopped forming stars at lower redshifts.
Galaxies below the critical mass continue to form stars at lower redshifts.

While the simple prescriptions that we have used for star formation and its 
shutdown are clearly only crude approximations, their success in 
simultaneously matching all the observed bimodality features at different 
redshifts indicates that they fairly represent the effects of the complex 
physical processes involved. 
This motivates a detailed study of the processes of cold-flow induced
star formation and the subsequent shutdown above the critical
mass at late times due to the coupling between virial shock heating and 
AGN feedback.


\section*{Acknowledgments}
We acknowledge stimulating discussions with J.~Binney, Y.~Birnboim,
S.M.~Faber, A.~Kravtsov, C.~Maulbetsch, E.~Neistein, J.R.~Primack, 
M.J.~Rees and R.~Somerville.
This research has been supported by ISF 213/02 and NASA ATP NAG5-8218.
A. Cattaneo has been supported by a Golda Meir Fellowship at HU.
A. Dekel acknowledges support from
a Miller Visiting Professorship at UC Berkeley,
a Visiting Professorship at UC Santa Cruz, and a Blaise Pascal International
Chair by the {\'E}cole Normale Sup{\'e}rieure at the Institut 
d'Astrophysique de Paris.


\bibliographystyle{mn2e}

\bibliography{ref_av}

\begin{thebibliography}{}

\bibitem[\protect\citeauthoryear{{Abazajian}, {Zheng}, {Zehavi}, {Weinberg},
  {Frieman}, {Berlind}, Blanton \& {et al.,}}{{Abazajian}
  et~al.}{2004}]{zehavi04_hod}
{Abazajian} K.,  {Zheng} Z.,  {Zehavi} I.,  {Weinberg} D.~H.,  {Frieman} J.~A.,
   {Berlind} A.~A.,  Blanton M.~R.,    {et al.,} 2004, astro-ph/0408003

\bibitem[\protect\citeauthoryear{{Baldry}, {Glazebrook}, {Brinkmann}, {Ivezi{\'
  c}}, {Lupton}, {Nichol} \& {Szalay}}{{Baldry} et~al.}{2004}]{baldry_etal04}
{Baldry} I.~K.,  {Glazebrook} K.,  {Brinkmann} J.,  {Ivezi{\' c}} {\v Z}.,
  {Lupton} R.~H.,  {Nichol} R.~C.,    {Szalay} A.~S.,  2004, \apj, 600, 681

\bibitem[\protect\citeauthoryear{{Balogh}, {Baldry}, {Nichol}, {Miller},
  {Bower} \& {Glazebrook}}{{Balogh} et~al.}{2004}]{balogh_etal04}
{Balogh} M.~L.,  {Baldry} I.~K.,  {Nichol} R.,  {Miller} C.,  {Bower} R.,
  {Glazebrook} K.,  2004, \apjl, 615, L101

\bibitem[\protect\citeauthoryear{{Begelman} \& {Nath}}{{Begelman} \&
  {Nath}}{2005}]{begelman_nath05}
{Begelman} M.~C.,  {Nath} B.~B.,  2005, \mnras, 361, 1387

\bibitem[\protect\citeauthoryear{{Bell}, {McIntosh}, {Katz} \&
  {Weinberg}}{{Bell} et~al.}{2003}]{bell_etal03}
{Bell} E.~F.,  {McIntosh} D.~H.,  {Katz} N.,    {Weinberg} M.~D.,  2003, \apjs,
  149, 289

\bibitem[\protect\citeauthoryear{{Bell}, {Wolf}, {Meisenheimer}, {Rix},
  {Borch}, {Dye}, {Kleinheinrich}, {Wisotzki} \& {McIntosh}}{{Bell}
  et~al.}{2004}]{bell_etal04}
{Bell} E.~F.,  {Wolf} C.,  {Meisenheimer} K.,  {Rix} H.,  {Borch} A.,  {Dye}
  S.,  {Kleinheinrich} M.,  {Wisotzki} L.,    {McIntosh} D.~H.,  2004, \apj,
  608, 752

\bibitem[\protect\citeauthoryear{{Binney}}{{Binney}}{1977}]{binney77}
{Binney} J.,  1977, \apj, 215, 483

\bibitem[\protect\citeauthoryear{{Binney}}{{Binney}}{2004}]{binney04}
{Binney} J.,  2004, \mnras, 347, 1093

\bibitem[\protect\citeauthoryear{{Binney} \& {Merrifield}}{{Binney} \&
  {Merrifield}}{1998}]{binney_merrifield98}
{Binney} J.,  {Merrifield} M.,  1998, {Galactic astronomy. Princeton Univ.
  Press, Princeton}

\bibitem[\protect\citeauthoryear{{Binney} \& {Tabor}}{{Binney} \&
  {Tabor}}{1995}]{binney_tabor95}
{Binney} J.,  {Tabor} G.,  1995, \mnras, 276, 663

\bibitem[\protect\citeauthoryear{{Birnboim} \& {Dekel}}{{Birnboim} \&
  {Dekel}}{2003}]{birnboim_dekel03}
{Birnboim} Y.,  {Dekel} A.,  2003, \mnras, 345, 349

\bibitem[\protect\citeauthoryear{{Blaizot}, {Guiderdoni}, {Devriendt},
  {Bouchet}, {Hatton} \& {Stoehr}}{{Blaizot} et~al.}{2004}]{blaizot_etal04}
{Blaizot} J.,  {Guiderdoni} B.,  {Devriendt} J.~E.~G.,  {Bouchet} F.~R.,
  {Hatton} S.~J.,    {Stoehr} F.,  2004, \mnras, 352, 571

\bibitem[\protect\citeauthoryear{{Blandford} \& {Begelman}}{{Blandford} \&
  {Begelman}}{1999}]{blandford_begelman99}
{Blandford} R.~D.,  {Begelman} M.~C.,  1999, \mnras, 303, L1

\bibitem[\protect\citeauthoryear{{Blanton}, {Eisenstein}, {Hogg} \&
  {Zehavi}}{{Blanton} et~al.}{2006}]{blanton_etal06}
{Blanton} M.~R.,  {Eisenstein} D.~J.,  {Hogg} D.~W.,    {Zehavi} I.,  2006,
  astro-ph/0411037

\bibitem[\protect\citeauthoryear{{Blumenthal}, {Faber}, {Primack} \&
  {Rees}}{{Blumenthal} et~al.}{1984}]{blumenthal_etal84}
{Blumenthal} G.~R.,  {Faber} S.~M.,  {Primack} J.~R.,    {Rees} M.~J.,  1984,
  \nat, 311, 517

\bibitem[\protect\citeauthoryear{{Br{\"u}ggen}, {Ruszkowski} \&
  {Hallman}}{{Br{\"u}ggen} et~al.}{2005}]{brueggen_etal05}
{Br{\"u}ggen} M.,  {Ruszkowski} M.,    {Hallman} E.,  2005, \apj, 630, 740

\bibitem[\protect\citeauthoryear{{Cattaneo} \& {Bernardi}}{{Cattaneo} \&
  {Bernardi}}{2003}]{cattaneo_bernardi03}
{Cattaneo} A.,  {Bernardi} M.,  2003, \mnras, 344, 45

\bibitem[\protect\citeauthoryear{{Cattaneo}, {Combes}, {Colombi}, {Bertin} \&
  {Melchior}}{{Cattaneo} et~al.}{2005}]{cattaneo_etal05}
{Cattaneo} A.,  {Combes} F.,  {Colombi} S.,  {Bertin} E.,    {Melchior} A.-L.,
  2005, \mnras, 359, 1237

\bibitem[\protect\citeauthoryear{{Chapman}, {Blain}, {Ivison} \&
  {Smail}}{{Chapman} et~al.}{2003}]{chapman_etal03}
{Chapman} S.~C.,  {Blain} A.~W.,  {Ivison} R.~J.,    {Smail} I.~R.,  2003,
  \nat, 422, 695

\bibitem[\protect\citeauthoryear{{Chapman}, {Smail}, {Blain} \&
  {Ivison}}{{Chapman} et~al.}{2004}]{chapman_etal04}
{Chapman} S.~C.,  {Smail} I.,  {Blain} A.~W.,    {Ivison} R.~J.,  2004, \apj,
  614, 671

\bibitem[\protect\citeauthoryear{{Ciotti} \& {Ostriker}}{{Ciotti} \&
  {Ostriker}}{1997}]{ciotti_ostriker97}
{Ciotti} L.,  {Ostriker} J.~P.,  1997, \apjl, 487, L105+

\bibitem[\protect\citeauthoryear{{Cole}, {Lacey}, {Baugh} \& {Frenk}}{{Cole}
  et~al.}{2000}]{cole_etal00}
{Cole} S.,  {Lacey} C.~G.,  {Baugh} C.~M.,    {Frenk} C.~S.,  2000, \mnras,
  319, 168

\bibitem[\protect\citeauthoryear{{Croton}, {Springel}, {White}, {De Lucia},
  {Frenk}, {Gao}, {Jenkins}, {Kauffmann}, {Navarro} \& {Yoshida}}{{Croton}
  et~al.}{2006}]{croton_etal06}
{Croton} D.~J.,  {Springel} V.,  {White} S.~D.~M.,  {De Lucia} G.,  {Frenk}
  C.~S.,  {Gao} L.,  {Jenkins} A.,  {Kauffmann} G.,  {Navarro} J.~F.,
  {Yoshida} N.,  2006, astro-ph/0508046

\bibitem[\protect\citeauthoryear{{Davis}, {Efstathiou}, {Frenk} \&
  {White}}{{Davis} et~al.}{1985}]{davis_etal85}
{Davis} M.,  {Efstathiou} G.,  {Frenk} C.~S.,    {White} S.~D.~M.,  1985, \apj,
  292, 371

\bibitem[\protect\citeauthoryear{{Dehnen} \& {Binney}}{{Dehnen} \&
  {Binney}}{1998}]{dehnen_binney98}
{Dehnen} W.,  {Binney} J.,  1998, \mnras, 294, 429

\bibitem[\protect\citeauthoryear{{Dekel} \& {Birnboim}}{{Dekel} \&
  {Birnboim}}{2006}]{dekel_birnboim06}
{Dekel} A.,  {Birnboim} Y.,  2006, astro-ph/0412300

\bibitem[\protect\citeauthoryear{{Dekel} \& {Silk}}{{Dekel} \&
  {Silk}}{1986}]{dekel_silk86}
{Dekel} A.,  {Silk} J.,  1986, \apj, 303, 39

\bibitem[\protect\citeauthoryear{{Devriendt}, {Guiderdoni} \&
  {Sadat}}{{Devriendt} et~al.}{1999}]{devriendt_etal99}
{Devriendt} J.~E.~G.,  {Guiderdoni} B.,    {Sadat} R.,  1999, \aap, 350, 381

\bibitem[\protect\citeauthoryear{{Di Matteo}, {Croft}, {Springel} \&
  {Hernquist}}{{Di Matteo} et~al.}{2003}]{dimatteo_etal03}
{Di Matteo} T.,  {Croft} R.~A.~C.,  {Springel} V.,    {Hernquist} L.,  2003,
  \apj, 593, 56

\bibitem[\protect\citeauthoryear{{Di Matteo}, {Springel} \& {Hernquist}}{{Di
  Matteo} et~al.}{2005}]{dimatteo_etal05}
{Di Matteo} T.,  {Springel} V.,    {Hernquist} L.,  2005, \nat, 433, 604

\bibitem[\protect\citeauthoryear{{Fabian}, {Reynolds}, {Taylor} \&
  {Dunn}}{{Fabian} et~al.}{2005}]{fabian_etal05}
{Fabian} A.~C.,  {Reynolds} C.~S.,  {Taylor} G.~B.,    {Dunn} R.~J.~H.,  2005,
  \mnras, 363, 891

\bibitem[\protect\citeauthoryear{{Fabian}, {Sanders}, {Allen}, {Crawford},
  {Iwasawa}, {Johnstone}, {Schmidt} \& {Taylor}}{{Fabian}
  et~al.}{2003}]{fabian_etal03}
{Fabian} A.~C.,  {Sanders} J.~S.,  {Allen} S.~W.,  {Crawford} C.~S.,  {Iwasawa}
  K.,  {Johnstone} R.~M.,  {Schmidt} R.~W.,    {Taylor} G.~B.,  2003, \mnras,
  344, L43

\bibitem[\protect\citeauthoryear{{Fardal}, {Katz}, {Gardner}, {Hernquist},
  {Weinberg} \& {Dav{\'e}}}{{Fardal} et~al.}{2001}]{fardal_etal01}
{Fardal} M.~A.,  {Katz} N.,  {Gardner} J.~P.,  {Hernquist} L.,  {Weinberg}
  D.~H.,    {Dav{\'e}} R.,  2001, \apj, 562, 605

\bibitem[\protect\citeauthoryear{{Giavalisco}, {Dickinson}, {Ferguson},
  {Ravindranath}, {Kretchmer}, {Moustakas}, {Madau}, {Fall}, {Gardner},
  {Livio}, {Papovich}, {Renzini}, {Spinrad}, {Stern} \& {Riess}}{{Giavalisco}
  et~al.}{2004}]{giavalisco_etal04}
{Giavalisco} M.,  {Dickinson} M.,  {Ferguson} H.~C.,  {Ravindranath} S.,
  {Kretchmer} C.,  {Moustakas} L.~A.,  {Madau} P.,  {Fall} S.~M.,  {Gardner}
  J.~P.,  {Livio} M.,  {Papovich} C.,  {Renzini} A.,  {Spinrad} H.,  {Stern}
  D.,    {Riess} A.,  2004, \apjl, 600, L103

\bibitem[\protect\citeauthoryear{{Granato}, {De Zotti}, {Silva}, {Bressan} \&
  {Danese}}{{Granato} et~al.}{2004}]{granato_etal04}
{Granato} G.~L.,  {De Zotti} G.,  {Silva} L.,  {Bressan} A.,    {Danese} L.,
  2004, \apj, 600, 580

\bibitem[\protect\citeauthoryear{{Granato}, {Silva}, {Monaco}, {Panuzzo},
  {Salucci}, {De Zotti} \& {Danese}}{{Granato} et~al.}{2001}]{granato_etal01}
{Granato} G.~L.,  {Silva} L.,  {Monaco} P.,  {Panuzzo} P.,  {Salucci} P.,  {De
  Zotti} G.,    {Danese} L.,  2001, \mnras, 324, 757

\bibitem[\protect\citeauthoryear{{Guiderdoni}, {Hivon}, {Bouchet} \&
  {Maffei}}{{Guiderdoni} et~al.}{1998}]{guiderdoni_etal98}
{Guiderdoni} B.,  {Hivon} E.,  {Bouchet} F.~R.,    {Maffei} B.,  1998, \mnras,
  295, 877

\bibitem[\protect\citeauthoryear{{H{\" a}ring} \& {Rix}}{{H{\" a}ring} \&
  {Rix}}{2004}]{haering_rix04}
{H{\" a}ring} N.,  {Rix} H.,  2004, \apjl, 604, L89

\bibitem[\protect\citeauthoryear{{Hatton}, {Devriendt}, {Ninin}, {Bouchet},
  {Guiderdoni} \& {Vibert}}{{Hatton} et~al.}{2003}]{hatton_etal03}
{Hatton} S.,  {Devriendt} J.~E.~G.,  {Ninin} S.,  {Bouchet} F.~R.,
  {Guiderdoni} B.,    {Vibert} D.,  2003, \mnras, 343, 75

\bibitem[\protect\citeauthoryear{{Helsdon} \& {Ponman}}{{Helsdon} \&
  {Ponman}}{2003}]{helsdon_ponman03}
{Helsdon} S.~F.,  {Ponman} T.~J.,  2003, \mnras, 340, 485

\bibitem[\protect\citeauthoryear{{Hernquist}}{{Hernquist}}{1990}]{Hernquist90}
{Hernquist} L.,  1990, \apj, 356, 359

\bibitem[\protect\citeauthoryear{{Hogg}, {Blanton}, {Brinchmann}, {Eisenstein},
  {Schlegel}, {Gunn}, {McKay}, {Rix}, {Bahcall}, {Brinkmann} \&
  {Meiksin}}{{Hogg} et~al.}{2004}]{hogg_etal04}
{Hogg} D.~W.,  {Blanton} M.~R.,  {Brinchmann} J.,  {Eisenstein} D.~J.,
  {Schlegel} D.~J.,  {Gunn} J.~E.,  {McKay} T.~A.,  {Rix} H.,  {Bahcall} N.~A.,
   {Brinkmann} J.,    {Meiksin} A.,  2004, \apjl, 601, L29

\bibitem[\protect\citeauthoryear{{Hubble}}{{Hubble}}{1926}]{hubble26}
{Hubble} E.~P.,  1926, \apj, 64, 321

\bibitem[\protect\citeauthoryear{{Humason}}{{Humason}}{1936}]{humason36}
{Humason} M.~L.,  1936, \apj, 83, 10

\bibitem[\protect\citeauthoryear{{Kannappan}}{{Kannappan}}{2004}]{kannappan04}
{Kannappan} S.~J.,  2004, \apjl, 611, L89

\bibitem[\protect\citeauthoryear{{Kauffmann}, {Heckman}, {White}, {Charlot},
  {Tremonti}, {Peng}, {Seibert}, {Brinkmann}, {Nichol}, {SubbaRao} \&
  {York}}{{Kauffmann} et~al.}{2003}]{kauffmann_etal03b}
{Kauffmann} G.,  {Heckman} T.~M.,  {White} S.~D.~M.,  {Charlot} S.,  {Tremonti}
  C.,  {Peng} E.~W.,  {Seibert} M.,  {Brinkmann} J.,  {Nichol} R.~C.,
  {SubbaRao} M.,    {York} D.,  2003, \mnras, 341, 54

\bibitem[\protect\citeauthoryear{{Kauffmann}, {White}, {Heckman}, {M{\'
  e}nard}, {Brinchmann}, {Charlot}, {Tremonti} \& {Brinkmann}}{{Kauffmann}
  et~al.}{2004}]{kauffmann_etal04}
{Kauffmann} G.,  {White} S.~D.~M.,  {Heckman} T.~M.,  {M{\' e}nard} B.,
  {Brinchmann} J.,  {Charlot} S.,  {Tremonti} C.,    {Brinkmann} J.,  2004,
  \mnras, 353, 713

\bibitem[\protect\citeauthoryear{{Kennicutt}}{{Kennicutt}}{1983}]{kennicutt83}
{Kennicutt} R.~C.,  1983, \apj, 272, 54

\bibitem[\protect\citeauthoryear{{Kere{\v s}}, {Katz}, {Weinberg} \&
  {Dav{\'e}}}{{Kere{\v s}} et~al.}{2005}]{keres_etal05}
{Kere{\v s}} D.,  {Katz} N.,  {Weinberg} D.~H.,    {Dav{\'e}} R.,  2005,
  \mnras, 363, 2

\bibitem[\protect\citeauthoryear{{Kravtsov}, {Berlind}, {Wechsler}, {Klypin},
  {Gottloeber}, {Allgood} \& {Primack}}{{Kravtsov} et~al.}{2004}]{krav04_hod}
{Kravtsov} A.~V.,  {Berlind} A.~A.,  {Wechsler} R.~H.,  {Klypin} A.~A.,
  {Gottloeber} S.,  {Allgood} B.,    {Primack} J.~R.,  2004, astro-ph/0308519

\bibitem[\protect\citeauthoryear{{Magorrian}, {Tremaine}, {Richstone},
  {Bender}, {Bower}, {Dressler}, {Faber}, {Gebhardt}, {Green}, {Grillmair},
  {Kormendy} \& {Lauer}}{{Magorrian} et~al.}{1998}]{magorrian_etal98}
{Magorrian} J.,  {Tremaine} S.,  {Richstone} D.,  {Bender} R.,  {Bower} G.,
  {Dressler} A.,  {Faber} S.~M.,  {Gebhardt} K.,  {Green} R.,  {Grillmair} C.,
  {Kormendy} J.,    {Lauer} T.,  1998, \aj, 115, 2285

\bibitem[\protect\citeauthoryear{{Marconi} \& {Hunt}}{{Marconi} \&
  {Hunt}}{2003}]{marconi_hunt03}
{Marconi} A.,  {Hunt} L.~K.,  2003, \apjl, 589, L21

\bibitem[\protect\citeauthoryear{{Mathews} \& {Brighenti}}{{Mathews} \&
  {Brighenti}}{2003}]{mathews_brighenti03}
{Mathews} W.~G.,  {Brighenti} F.,  2003, \araa, 41, 191

\bibitem[\protect\citeauthoryear{{McKee} \& {Ostriker}}{{McKee} \&
  {Ostriker}}{1977}]{mckee_ostriker77}
{McKee} C.~F.,  {Ostriker} J.~P.,  1977, \apj, 218, 148

\bibitem[\protect\citeauthoryear{{McNamara}, {Nulsen}, {Wise}, {Rafferty},
  {Carilli}, {Sarazin} \& {Blanton}}{{McNamara} et~al.}{2005}]{mcnamara_etal05}
{McNamara} B.~R.,  {Nulsen} P.~E.~J.,  {Wise} M.~W.,  {Rafferty} D.~A.,
  {Carilli} C.,  {Sarazin} C.~L.,    {Blanton} E.~L.,  2005, \nat, 433, 45

\bibitem[\protect\citeauthoryear{{Merritt} \& {Ferrarese}}{{Merritt} \&
  {Ferrarese}}{2001}]{merritt_ferrarese01}
{Merritt} D.,  {Ferrarese} L.,  2001, \apj, 547, 140

\bibitem[\protect\citeauthoryear{{Moustakas}, {Casertano}, {Conselice},
  {Dickinson}, {Eisenhardt}, {Ferguson}, {Giavalisco}, {Grogin}, {Koekemoer},
  {Lucas}, {Mobasher}, {Papovich}, {Renzini}, {Somerville} \&
  {Stern}}{{Moustakas} et~al.}{2004}]{moustakas_etal04}
{Moustakas} L.~A.,  {Casertano} S.,  {Conselice} C.~J.,  {Dickinson} M.~E.,
  {Eisenhardt} P.,  {Ferguson} H.~C.,  {Giavalisco} M.,  {Grogin} N.~A.,
  {Koekemoer} A.~M.,  {Lucas} R.~A.,  {Mobasher} B.,  {Papovich} C.,  {Renzini}
  A.,  {Somerville} R.~S.,    {Stern} D.,  2004, \apjl, 600, L131

\bibitem[\protect\citeauthoryear{{Nulsen} \& {Fabian}}{{Nulsen} \&
  {Fabian}}{2000}]{nulsen_fabian00}
{Nulsen} P.~E.~J.,  {Fabian} A.~C.,  2000, \mnras, 311, 346

\bibitem[\protect\citeauthoryear{{Omma}, {Binney}, {Bryan} \& {Slyz}}{{Omma}
  et~al.}{2004}]{omma_etal04}
{Omma} H.,  {Binney} J.,  {Bryan} G.,    {Slyz} A.,  2004, \mnras, 348, 1105

\bibitem[\protect\citeauthoryear{{Osmond} \& {Ponman}}{{Osmond} \&
  {Ponman}}{2004}]{osmond_ponman04}
{Osmond} J.~P.~F.,  {Ponman} T.~J.,  2004, \mnras, 350, 1511

\bibitem[\protect\citeauthoryear{{Rees} \& {Ostriker}}{{Rees} \&
  {Ostriker}}{1977}]{rees_ostriker77}
{Rees} M.~J.,  {Ostriker} J.~P.,  1977, \mnras, 179, 541

\bibitem[\protect\citeauthoryear{{Reynolds}, {Heinz} \& {Begelman}}{{Reynolds}
  et~al.}{2001}]{reynolds_etal01}
{Reynolds} C.~S.,  {Heinz} S.,    {Begelman} M.~C.,  2001, \apjl, 549, L179

\bibitem[\protect\citeauthoryear{{Ruszkowski}, {Br{\"u}ggen} \&
  {Begelman}}{{Ruszkowski} et~al.}{2004}]{ruszkowski_etal04}
{Ruszkowski} M.,  {Br{\"u}ggen} M.,    {Begelman} M.~C.,  2004, \apj, 615, 675

\bibitem[\protect\citeauthoryear{{Sawicki} \& {Thompson}}{{Sawicki} \&
  {Thompson}}{2005}]{sawicki_thompson05}
{Sawicki} M.,  {Thompson} D.,  2005, \apj, 635, 100

\bibitem[\protect\citeauthoryear{{Shapley}, {Erb}, {Pettini}, {Steidel} \&
  {Adelberger}}{{Shapley} et~al.}{2004}]{shapley_etal04}
{Shapley} A.~E.,  {Erb} D.~K.,  {Pettini} M.,  {Steidel} C.~C.,    {Adelberger}
  K.~L.,  2004, \apj, 612, 108

\bibitem[\protect\citeauthoryear{{Silk}}{{Silk}}{1977}]{silk77}
{Silk} J.,  1977, \apj, 211, 638

\bibitem[\protect\citeauthoryear{{Silk} \& {Rees}}{{Silk} \&
  {Rees}}{1998}]{silk_rees98}
{Silk} J.,  {Rees} M.~J.,  1998, \aap, 331, L1

\bibitem[\protect\citeauthoryear{{Smail}, {Ivison}, {Blain} \& {Kneib}}{{Smail}
  et~al.}{2002}]{smail_etal02}
{Smail} I.,  {Ivison} R.~J.,  {Blain} A.~W.,    {Kneib} J.-P.,  2002, \mnras,
  331, 495

\bibitem[\protect\citeauthoryear{{Somerville} \& {Primack}}{{Somerville} \&
  {Primack}}{1999}]{somerville_primack99}
{Somerville} R.~S.,  {Primack} J.~R.,  1999, \mnras, 310, 1087

\bibitem[\protect\citeauthoryear{{Springel}, {Di Matteo} \&
  {Hernquist}}{{Springel} et~al.}{2005}]{springel_etal05}
{Springel} V.,  {Di Matteo} T.,    {Hernquist} L.,  2005, \apjl, 620, L79

\bibitem[\protect\citeauthoryear{{Steidel}, {Adelberger}, {Giavalisco},
  {Dickinson} \& {Pettini}}{{Steidel} et~al.}{1999}]{steidel_etal99}
{Steidel} C.~C.,  {Adelberger} K.~L.,  {Giavalisco} M.,  {Dickinson} M.,
  {Pettini} M.,  1999, \apj, 519, 1

\bibitem[\protect\citeauthoryear{{Stockton}}{{Stockton}}{1999}]{stockton99}
{Stockton} A.,  1999, in {IAU Symp. 186: Galaxy Interactions at Low and High
  Redshift} p.~311

\bibitem[\protect\citeauthoryear{{Sutherland} \& {Dopita}}{{Sutherland} \&
  {Dopita}}{1993}]{sutherland_dopita93}
{Sutherland} R.~S.,  {Dopita} M.~A.,  1993, \apjs, 88, 253

\bibitem[\protect\citeauthoryear{{Tabor} \& {Binney}}{{Tabor} \&
  {Binney}}{1993}]{tabor_binney93}
{Tabor} G.,  {Binney} J.,  1993, \mnras, 263, 323

\bibitem[\protect\citeauthoryear{{Toomre} \& {Toomre}}{{Toomre} \&
  {Toomre}}{1972}]{toomre_toomre72}
{Toomre} A.,  {Toomre} J.,  1972, \apj, 178, 623

\bibitem[\protect\citeauthoryear{{Tremaine}, {Gebhardt}, {Bender}, {Bower},
  {Dressler}, {Faber}, {Filippenko}, {Green}, {Grillmair}, {Ho}, {Kormendy},
  {Lauer}, {Magorrian}, {Pinkney} \& {Richstone}}{{Tremaine}
  et~al.}{2002}]{tremaine_etal02}
{Tremaine} S.,  {Gebhardt} K.,  {Bender} R.,  {Bower} G.,  {Dressler} A.,
  {Faber} S.~M.,  {Filippenko} A.~V.,  {Green} R.,  {Grillmair} C.,  {Ho}
  L.~C.,  {Kormendy} J.,  {Lauer} T.~R.,  {Magorrian} J.,  {Pinkney} J.,
  {Richstone} D.,  2002, \apj, 574, 740

\bibitem[\protect\citeauthoryear{{Tucker} \& {David}}{{Tucker} \&
  {David}}{1997}]{tucker_david97}
{Tucker} W.,  {David} L.~P.,  1997, \apj, 484, 602

\bibitem[\protect\citeauthoryear{{van den Bosch}}{{van den
  Bosch}}{1998}]{vandenbosch98}
{van den Bosch} F.~C.,  1998, \apj, 507, 601

\bibitem[\protect\citeauthoryear{{van der Marel}}{{van der
  Marel}}{}]{vandermarel99}
{van der Marel} R.~P.,

\bibitem[\protect\citeauthoryear{{White} \& {Frenk}}{{White} \&
  {Frenk}}{1991}]{white_frenk91}
{White} S.~D.~M.,  {Frenk} C.~S.,  1991, \apj, 379, 52

\bibitem[\protect\citeauthoryear{{White} \& {Rees}}{{White} \&
  {Rees}}{1978}]{white_rees78}
{White} S.~D.~M.,  {Rees} M.~J.,  1978, \mnras, 183, 341

\bibitem[\protect\citeauthoryear{{Wisotzki}, {Kuhlbrodt} \&
  {Jahnke}}{{Wisotzki} et~al.}{2001}]{wisotzki_etal01}
{Wisotzki} L.,  {Kuhlbrodt} B.,    {Jahnke} K.,  2001, in QSO Hosts and Their
  Environments p.~83

\bibitem[\protect\citeauthoryear{{Yan}, {Madgwick} \& {White}}{{Yan}
  et~al.}{2003}]{yan03_hod_2df}
{Yan} R.,  {Madgwick} D.~S.,    {White} M.,  2003, \apj, 598, 848

\end{thebibliography}

\label{lastpage}
\end{document}